\documentclass[11pt,3p,times,twoside,sort,espcrc2,preprint]{elsarticle}

\makeatletter
\def\ps@pprintTitle{%
 \let\@oddhead\@empty
 \let\@evenhead\@empty 
 \def\@oddfoot{\centerline{\thepage}}%
 \let\@evenfoot\@oddfoot}
\makeatother

\usepackage{graphicx}  % needed for figures
\usepackage{dcolumn}   % needed for some tables
\usepackage{bm}        % for math
\usepackage{amssymb}   % for math
\usepackage{textcomp}
\usepackage{mathdots}
\usepackage{booktabs}
\usepackage{color}
\usepackage{amsmath}
\usepackage{amssymb}
\usepackage{bm,braket}
\usepackage{lipsum}
\usepackage{xargs} 
\usepackage{hyperref}
\usepackage[pdftex,dvipsnames]{xcolor}  % Coloured text etc.
\usepackage{siunitx}
\usepackage[outdir=./]{epstopdf}
\DeclareSIUnit \pc {pc}
\DeclareSIUnit \Mpc {Mpc}
\DeclareSIUnit \MB {MB}
\DeclareSIUnit \GB {GB}
\DeclareSIUnit \TB {TB}

\newcommand\numberthis{\addtocounter{equation}{1}\tag{\theequation}}
\newcommand{\mbeq}{\overset{!}{=}}

\newcommand{\bea}{\begin{eqnarray}}
\newcommand{\eea}{\end{eqnarray}}
\newcommand{\be}{\begin{equation}}
\newcommand{\ee}{\end{equation}}

\newcommand{\rev}[1]{\textcolor{black}{#1}}
\newcommand{\edit}[1]{\textcolor{black}{#1}}

\DeclareMathOperator{\const}{const.}

\journal{Computer Physics Communications}

\begin{document}

\begin{frontmatter}

\title{%
    Comparison of two different integration methods for the (1+1)-Dimensional 
    Schr\"odinger-Poisson Equation
}

\author[unihd]{Nico Schwersenz,\corref{dlr}}
\author[univa]{Victor Loaiza}
\author[uio]{Tim Zimmermann}
\author[univa]{Javier~Madro\~nero}
\author[unipa,unicl]{Sandro Wimberger,\corref{cor}}
\cortext[dlr]{Current address: Deutsches Zentrum f\"ur Luft- und Raumfahrt e.V. (DLR), Wilhelm-Runge-Stra{\ss}e 10, 89081 Ulm, Germany}
\cortext[cor]{Corresponding author, e-mail: sandromarcel.wimberger@unipr.it}
\address[unihd]{%
    Institut f\"ur Theoretische Physik, Philosophenweg 16, Universit\"at Heidelberg, 69120 Heidelberg, Germany
}
\address[univa]{Centre for Bioinformatics and Photonics (CIBioFi), Universidad del Valle, Edificio E20, No. 1069, 760032 Cali, Colombia
}
\address[uio]{%
   Institute of Theoretical Astrophysics, University of Oslo, Postboks 1029, 0315 Oslo, Norway 
}
\address[unipa]{%
    Dipartimento di Scienze Matematiche, Fisiche e Informatiche, Universit\`a di Parma,
    Campus Universitario, Parco Area delle Scienze n. 7/a, 43124 Parma, Italy
}
\address[unicl]{INFN, Sezione di Milano Bicocca, Gruppo Collegato di Parma, Italy}

\begin{abstract}
We compare two different numerical methods to integrate in time spatially delocalized initial \rev{densities using the Schr\"odinger-Poisson equation system as the evolution law.} 
The basic equation is a nonlinear Schr\"odinger equation with an auto-gravitating potential created by the wave function density itself.
The latter is determined as a solution of Poisson's equation modelling\rev{, e.g.,} non-relativistic gravity. For reasons of complexity, we
treat a one-dimensional version of the problem whose numerical integration is still challenging because of the extreme long-range 
forces (being constant in the asymptotic limit). Both of our methods, a Strang splitting scheme and  a basis function approach 
using B-splines, are compared in numerical convergence and effectivity. Overall, our Strang-splitting evolution compares favourably with the B-spline method. 
In particular, by using an adaptive time-stepper rather large one-dimensional boxes can be treated. These results
give hope for extensions to two spatial dimensions for not too small boxes and large evolution times necessary for describing\rev{, for instance,}
 dark matter formation over cosmologically relevant scales.
\end{abstract}

\end{frontmatter}

\section{Introduction}
\label{sec:intro}

Nonlinear Schr\"odinger equations (NLSE) with (non-)local interactions constitute a
physically intriguing and phenomenologically rich model for a range of seemingly
disparate systems. If the interaction is local and its coupling strength
constant in time, the Gross-Pitaevskii equation represents an accurate
mean-field description of Bose-Einstein condensates in $d\leq3$
dimensions \cite{Pitaevskii2016}. Experimentally observable phenomena such as superfluidity or
quantum vortices are theoretically recovered within the scope of this NLSE.
If the interaction is long-range and identical to the Green's
function of Poisson's equation, the Schr\"odinger-Poisson equation (SP) with
\emph{constant} coupling strength --- a NLSE --- is used in nonlinear optics as $d<3$ 
steady-state model of light propagating through dispersive media
\cite{Picozzi2011, Bekenstein2015, Roger2016, Navarrete2017}.
In cosmology, one recovers SP with \emph{time-dependent} interaction strength as the non-relativistic 
equation of motion of a scalar field minimally coupled to space-time geometry in
$d=3$ spatial dimensions, e.g. \cite{Chavanis2012}. This model is known in cosmology as 
fuzzy dark matter \cite{Hu2000}.
In both cases, the Gross-Pitaevskii and the SP equation, 
nonlinear effects such as soliton formation \cite{Guzman2006,Levkov2018},
evolution \cite{Li2021}, and interaction with transient interference fringes are accurately predicted by the
model, e.g. \cite{Li2021,Zagorac2023}. 
Fuzzy dark matter provides an alternative approach in order to model large scale structure formation by the evolution of a complex scalar field, the
wave function in the SP system.  
\rev{Using only the dark matter without the baryonic part is certainly an approximation for the real universe, but for some purposes such as the formation of structures on large, intergalactic scales, it
is believed to be nevertheless realistic because of the dominance of dark matter with respect to the baryonic matter \cite{Peebles1980}.} 
Depending on one's viewpoint, one may either interpret such a description of dark matter as structure formation driven by a cosmic 
Bose--Einstein condensate trapped in its own gravitational potential \cite{Hu2000, Hui2017} and where the cosmic expansion determines the time-dependent interaction strength, 
or as a field theoretical approximation to the classical collisionless Boltzmann equation \cite{Kopp2017, Mocz2018}. In the latter sense, SP simulations may be seen as an 
alternative method to sample the phase space evolution of cold dark matter \cite{Kopp2017, Mocz2018, Bullock2017}. 
%%% Mostly mentioned is the fact that Fuzzy dark matter could provide a solution to the small scale crisis of other models of cold dark matter \cite{Bullock2017} without calling for sophisticated baryonic feedback processes.
\rev{The SP system finds its application in many other contexts such as in nonlinear optics \cite{Bekenstein2015, Kivshar2005} and in investigation
of structure formation on small scales in exotic Bose-Einstein condensates, see e.g. \cite{Akulin2000, Choi2002, Yakimenko2023}. The recent
review by Paredes et al. \cite{Paredes2020} conveniently collects the different physical motivations for the SP system.}

From a numerical point of view, integrating the three dimensional SP system for realistic cosmological initial conditions
poses a challenging problem. Its Hamiltonian is nonlinear and the nonlocal interaction leads to
strong mode coupling, forcing one to resolve all spatial scales down
to the microscopic de Broglie scale, below which Heisenberg's uncertainty
principle effectively implements a diffusion mechanism, see e.g. \cite{PRD2021}. 
Allowing for a time-dependent coupling strength complicates the formulation of a time-evolution operator and deprives us of energy
conservation already at the analytical level.
Multiple numerical approaches and semi-analytical approximations have emerged to
(partially) combat these issues in a top-down fashion, i.e. simplifications
applied to the full-fledged, three-dimensional problem:
Ref. \cite{Schive2014, Mina2020} use adaptive-mesh refinement techniques to alleviate
the high cost of resolving many orders of magnitude in spatial scale --- in
cosmological applications usually five to six orders --- at the expense of non-unitary time-evolution.
The authors of \cite{Nori2018} recast SP to an approximately correct hydrodynamic form and
apply smooth particle hydrodynamics to discretize and evolve the resulting
equations. This increases the computational efficiency substantially and makes
norm conservation manifest. However, due to the approximate nature of the
hydrodynamic formalism, one loses certain aspects of the full-fledged SP
phenomenology and interference effects in particular. Ref. \cite{Schwabe2022}
introduced a promising hybrid method in which the large scale dynamics is
followed by means of a phase-augmented $N$-body problem with which the SP
wave function can be reconstructed via a WKB-like gaussian beam decomposition in small scale, 
high density regions. This is ideal for the purpose of deep zoom in simulations
of highly compact objects. Application to cosmological box sizes is still
pending. The most widely used technique for integration of SP in three
dimensions is the pseudo-spectral or Fourier method,
\cite{Mocz2017,May2021,May2022}, cf. Sec. \ref{sec:PS}. The
method is second-order accurate in time and spectrally accurate in space.
Moreover, its structure lends itself to an easy implementation in existing astrophysical codes.
However, its major bottleneck is the requirement of an uniform spatial grid
which translates to run times of $\mathcal{O}(10^7)$ CPU hours for small to
moderately sized cosmological boxes \cite{May2021}.

The purpose of this work is to make progress on the bottom-up approach, i.e.
instead of tackling the $d=3$ SP system directly, we build a comprehensive
numerical understanding in lower dimensions, in this work $d=1$, first. In a
previous work, see ref. \cite{PRD2021}, we focused on the physical implications of this
approach and showed that by introducing an external confinement, important
features of the three-dimensional setting, in particular the correct
interaction range, can be implemented with only one spatial degree of freedom.
\rev{A one-dimensional setting finds its own motivation also in the possibility to scan 
broad parameter spaces and to allow for generating larger ensembles of simulations
with random initial conditions \cite{PRD2021}.}

In this work, we focus more on numerical aspects of the bottom-up
approach and integrate the $d=1$ SP system without confinement --- a
setting numerically more challenging than the situation with confinement due to the stronger
mode-coupling. Compared to the discussion in \cite{PRD2021}, we complicate the
task at hand even further by allowing for a time-dependent coupling strength in
all our considerations that models the expansion of the universe during the evolution.
Our main goal is to compare two numerical approaches, i.e. the aforementioned
pseudo-spectral method and a B-spline approach commonly used in atomic and
molecular physics, see e.g. \cite{Bachau_2001, Bspline-1, Bspline-3, Bspline-4}.  
Our results are: first, we are fairly sure that our numerical approach is well converged.
This is underpinned by a direct comparison of wave functions and by testing a series of conservation laws.
Second, we demonstrate that the pseudo-spectral method, in particular 
together with an advanced adaptive-time stepper, is most efficient for treating larger system sizes.
These results give us hope for an extension of a fully converged method to higher dimensions.

The paper is organized as follows: Sec.
\ref{sec:theory} gives details on the exact initial boundary value problem we
want to solve and summarizes key properties of the one-dimensional SP system.
Sec. \ref{sec:numerical_method} outlines our numerical methods. An in-depth
comparison between both the pseudo-spectral and B-spline method in terms of
accuracy and performance is given in Sec. \ref{sec:res}. 
Building on these results, we introduce a time-adaptive extension to the plain
pseudo-spectral approach and showcase its effectiveness in Sec.
\ref{sec:adaptive_stepper}. We conclude in Sec. \ref{sec:conclusion}.

\section{Theoretical background}
\label{sec:theory}

We begin with a compact discussion of the SP system in one
spatial dimension. As noted in our introductory remarks, SP arises as
challenging equation of motion in various contexts \rev{\cite{Paredes2020}}. Here, we are primarily
interested in SP at a numerical level. However, to make the problem of integrating SP
well-defined and sufficiently challenging, initial conditions, boundary conditions and 
parameters should be chosen in a physically reasonable manner. 
To do so, we take inspiration from cosmology where the three-dimensional SP
system arises as \rev{a dark-matter only approximation in} the non-relativistic 
limit of a scalar field coupled to an expanding background cosmology, \cite{Chavanis2012}. 
In this application the "wave function" models a complex scalar, but classical,
field. It's density, $|\psi|^2$, is interpreted as the density of dark matter
that evolves self-consistently in it's own gravitational potential \cite{Schive2014}. 

Section \ref{sub:SPE} introduces the SP system in one spatial dimension. 
In Sec. \ref{sub:ic} we elaborate on our choice of initial conditions, while Sec. \ref{sub:conservation_laws}
summarises important conservation properties of SP, which we use in Sec. \ref{sec:res} to benchmark 
our numerical results.

\subsection{\label{sub:SPE} 
    Schr\"odinger-Poisson equation
}

The equation to solve is the one-dimensional Schr\"odinger-Poisson equation or  
$(1+1)$-SP, where $+1$ stands for the time variable. It can be derived from the 
three-dimensional version by assuming a homogeneous matter distribution in the two 
transverse dimensions and factorizing the wave function
into the transverse directions and the direction of interest \cite{PRD2021, timmaster}.
In momentum ($k$) space, the remaining interaction kernel is recognisable as a lower 
dimensional analogue to the three-dimensional gravity kernel. 
In both cases, the spectrum scales as $1/k^2$. 
The $x$-space kernel, on the other hand, looks qualitatively different leading asymptotically to a 
constant force or a linear potential satisfying free-space boundary conditions. 
As analysed in detail in \cite{PRD2021}, the latter lacks a length scale which we 
can associate with a finite interaction range. 

Not only does this introduce distinct differences in the asymptotic system state 
and the underlying relaxation mechanism of the one-dimensional dynamics compared to the 
full-fledged three-dimensional situation.
The absence of a clear length scale complicates the numerical problem quite a bit, see e.g. \cite{Tanaka2017}. 
We note in passing that a similar behaviour -- the evolution towards nonequilibrium quasi-stationary states and extremely long 
relaxation times to equilibrium -- is well known from other long-range interacting classical systems, also in 
one-dimension as in our case here, see for instance \cite{Ruffo2002, Levin2013, Campa2014, Lapo2017}.

The coupled equations we evolve in time read in their dimensionless form in the co-moving frame:
\begin{align}
    i\partial_t\psi(x,t) &= \hat{H}(t) \psi(x,t) 
    = \left[\hat{H}_K + \hat{H}_V(t)\right] \psi(x,t) 
    = \left[-\frac{1}{2}\partial^2_x + a(t)V(|\psi(x,t)|^2)\right]
                            \psi(x,t)
    \label{eq:SP1} 
    \\
    \partial_x^2 V(|\psi(x,t)|^2) &= |\psi(x,t)|^2 - 1
    \label{eq:SP2} \;,
\end{align}
where $V$ is the nonlocal, nonlinear potential given by the convolution of the
density fluctuations $\delta \rho(x,t) \equiv |\psi(x,t)|^2 - 1$. Here we subtract the
uniform background since, in the co-moving frame, we are only interested in the density evolution induced 
by the initial cosmological density contrast, see e.g. \cite{Woo2009, Chavanis2012, Schive2014, Mocz2017, May2021, May2022}.
Effectively, we work in a finite box of length $L$, and we apply periodic boundary 
conditions:
\begin{equation}
    \begin{split}
    \label{eq:pbc}
    \psi(0,t) = \psi(L,t), \\
    \partial_x \psi(0,t) = \partial_x \psi(L,t) \;.
    \end{split}
\end{equation}
Then, the norm conservation of the wave function corresponds to the 
conservation of mass during the evolution:
\begin{equation}
\label{eq:norm}
M = \int_{0}^{L} |\psi(x, t)|^2 \,dx \mbeq L .
\end{equation}

Note that the enforced equality in Eq. \eqref{eq:norm} corresponds to a
compatibility condition for Poisson's equation, Eq. \eqref{eq:SP2}, with our
choice of periodic boundary condition: If $L\neq M$ then Eq. \eqref{eq:SP2}, together
with boundary conditions Eq. \eqref{eq:pbc}, would be ill-defined.
Moreover, since Eq. \eqref{eq:SP1} is nonlinear, it is not possible to satisfy Eq. 
\eqref{eq:norm} by normalizing an obtained numerical solution a posteriori. 
Hence a norm conserving, unitary time evolution is a necessity for the numerical 
treatment of Eq. \eqref{eq:SP1} and the simplest check of the quality of the numerics 
will be to compute the total norm or mass from the evolved wave function. 

A physically more insightful representation of the interaction potential
$V(|\psi(x,t|^2)$ follows from the integral solution of Eq. \eqref{eq:SP2}. Let
$G(x,x')$ denote the interaction kernel, i.e. the gravitational potential of a
point source at $x'$ experienced at $x$. We then have:
\be
    \label{eq:potential}
    V(x,t) \equiv V(|\psi(x,t)|^2) = \int_0^L dx' G(x,x')|\psi(x',t)|^2 \;.
\ee
$G(x,x')$ can be given analytically \cite{Marshall_2000, Barton1989}:
\be
    G(x,x') = \frac{1}{2} |x-x'| - \frac{1}{2}
    \left(\frac{(x-x')^2}{L} + \frac{L}{6}\right)
    = \frac{1}{L} \sum_{l>0} \left(-\frac{1}{k_l^2}\right) 
    \exp{(-ik_l(x-x'))}.
    \label{eq:kernel}
\ee
Here $k_l = \frac{2\pi l}{L}$, $l\in \{0,...,N-1\}$, for a spatial discretisation with $N$ grid points.
We absorbed the aforementioned compatibility condition into $G(x,x')$ by
suppressing the singular $k=0$ ($l=0$) mode in its spectrum.
This explains why the density $\rho(x,t) \equiv |\psi(x,t)|^2$ and not just the density contrast $\delta\rho(x,t)$ enters into the
convolution of Eq. \eqref{eq:potential}. Consequently, the spatial mean potential is zero, $\langle V \rangle_L=0$, 
which complies with the imposed periodicity in the spatial domain.

The time-dependent coupling $a(t) \in (0,1]$ --- in the cosmological context known as
scale factor modelling the expansion of the universe --- encapsulates our choice of background cosmology and is given by the solution of 
Friedmann's equation in a flat and radiation-free universe, see e.g. \cite{Ryden}:
\be
    \left(\frac{\Dot{a}}{a}\right)^2 = 
    \mathcal{H}^2_0 \left(\Omega_{m,0} a^{-3} 
    + \Omega_{\Lambda,0}\right)
    \label{eq:cosmology}.
\ee
Here the present-day density parameters $\Omega_{m,0}=0.3$ 
for the dark matter (\rev{as usual for a dark-matter only approximation} absorbing all 
background matter density into this single parameter \cite{Woo2009, May2021, May2022}) and 
$\Omega_{\Lambda,0}=0.7$ for the dark energy are used.
For today's Hubble constant we take the specific value
$\mathcal{H}_0=\SI{68}{\kilo\meter\per \second\Mpc}$ for our simulations, thus
making our choice of cosmological parameters, i.e.
$\{\Omega_{m,0}, \Omega_{\Lambda,0}, \mathcal{H}_0\}$,
compatible with observational constraints, see e.g. \cite{Planck2018}.
The reader is referred to \cite{Zimmermann_2019} for an illustration of the
temporal behavior of the scale factor. Under realistic conditions, the
integration is initialised with $a(t_0) \approx 10^{-2}$. Subsequently, $a(t)$ is 
strictly monotonically increasing and two distinct regimes may be identified.
For $t<40$, the background expansion, measured by the value $a(t)$, proceeds slowly so
that Eq. \eqref{eq:SP1} is close to free. Once $t>40$, $a(t)$ grows rapidly in a power-law like manner
thus increasing the importance of the nonlinear interaction term on a short time scale.

Section \ref{sec:res} will report results obtained by numerical simulations of Eqs. \eqref{eq:SP1},  \eqref{eq:SP2} and \eqref{eq:cosmology} 
with two different numerical methods.

\subsection{\label{sub:ic} Initial Conditions}

In \cite{PRD2021} most of our analysis focused on spatially localised,
\emph{synthetic} initial conditions, e.g. a gravitationally unstable gaussian wave packet.
The aforementioned nonlocality of the interaction kernel $G(x,x')$, on the
other hand, suggests that spatially delocalized initial conditions are more challenging to integrate
correctly.

In what follows, all simulations depart from \emph{cosmological} initial
conditions adapted to the one dimensional situation at hand. On the numerical
level relevant for this work, these may be understood as a random and spatially
delocalised realisation of the wave function where prior expectation of the
properties of $|\psi|^2$ are absorbed into a long-range spatial correlation. \rev{The long-range correlations, 
which dynamically manifest by the long-range potential, together with the double time-dependence of the potential through the varying
density as well as the fast varying scale factor $a(t)$, make the numerical problem challenging.}
To keep the discussion compact, we refer to \cite{timmaster} for a detailed 
exposition and only mention key points of the initial condition construction. 

Initialising the wave function at $t_0=0$ requires us to specify two degrees of freedom at
each grid location $x_n$ --- wave modulus $\delta(x_n,t_0)$ and phase $S(x_n, t_0)$:
\begin{equation}
    \psi(x_n, t_0) = \sqrt{\delta(x_n, t_0) + 1} \exp\left(i S(x_n, t_0)\right) \;.
\end{equation}
We may already deduce from this ansatz that the spatially averaged density contrast 
$\langle \delta \rangle_L = 0$ in order to satisfy Eq. \eqref{eq:norm} at the initial time $t_0$.

The modulus, is expected to obey gaussian statistics at early ($a(t) < 10^{-2}$) times \cite{Dodelson}. 
More precisely, if $\Sigma_{nm} = \xi(x_n, x_m; t_0) = \langle \delta(x_n, t_0)\delta(x_m, t_0)\rangle$ 
denotes the real space covariance matrix, then
$\bm\delta(x_n, t_0) ~ \sim~\mathcal{N}\left(0, \Sigma\right)$, for the discretized situation at hand.
$\mathcal{N}\left(0, \Sigma\right)$ denotes the normal distribution with zero mean and standard deviation $\Sigma$.
Note that the zero mean is a consequence of ergodicity so that
$\langle\delta\rangle = \langle \delta\rangle_L = 0$ by construction.

For long-range correlations, $\Sigma$ will be a
dense covariance matrix and sampling from this high dimension normal distribution
is thus intractable in the case of high resolution grids with $N = \mathcal{O}({10^6})$ grid
points. Fortunately, the cosmological principle, i.e. the assumption of statistical isotropy and
homogeneity, \cite{Peebles1980}, guarantee the one-dimensional matter power spectrum
$P(k; t_0) = \mathcal{F}[\xi(r; t_0)]$ to be diagonal in momentum space. It is
for this reason that the construction of cosmological initial conditions is usually carried out in $k$-space.

Form and time dependence of the one dimensional matter power spectrum is set by (i) the linear theory of
structure formation, \cite{Dodelson}, governing the three dimensional matter power spectrum $P_{3D}(k; t_0)$, and (ii) a
dimensional reduction from $P_{3D}(k; t_0)$ to $P(k; t_0)$. For the purpose of this work, we treat step (i) as black box and rely on 
well-established linear Boltzmann codes such as \texttt{AxionCAMB}, see \cite{Hlozek2017}, to provide the three dimensional matter
power spectrum at simulation start $t_0$. With $P_{3D}(k; t_0)$ at hand, we
proceed as in \cite{PRD2021, Hu2000} and find its lower dimensional analogue $P(k; t_0)$
by demanding a dimension independent variance $\xi_{3D}(0; t_0) = \xi(0; t_0)$.

Finally, the initial phase function $S(x_n, t_0)$ follows from mass conservation in the linear evolution regime
\cite{timmaster}:
\begin{equation}
    \partial^2 S(x, t_0) = - m a(t_0) \dot a(t_0) \delta(x, a(t_0)) \;,
\end{equation}
which we recognize as Poisson-type equation and thus solvable with the numerical
procedure outlined in Sec. \ref{sec:numerical_method} for the gravitational potential.

\subsection{\label{sub:conservation_laws}%
    Conserved Quantities of \texorpdfstring{$(1+1)$}{(1+1)}-SP
}

Lacking analytical solutions, a crucial part of our numerical test harness,
Sec. \ref{sec:res}, is to assess deviations of numerically deduced observables, 
such as mass, momentum and energy, from their analytical counterparts which are manifestly 
conserved in the dynamics described by Eq. \eqref{eq:SP1}. Here we summarize these conserved quantities.

Note that Eq. \eqref{eq:SP1} -  \eqref{eq:SP2} is the Euler-Lagrange equation of the action:
\begin{equation}
    \label{eq:action}
    \mathcal{S} = \int \text{d}t \int \text{d}x \left(i\psi\partial_t\psi -
    \mathcal{H}\right),
\end{equation}
with Hamiltonian density:
\begin{equation}
    \mathcal H 
    = 
    \frac{1}{2} \left\{|\partial_x \psi|^2 
        + a(t) \left(\int \text{d}x' G(x,x') |\psi(x')|^2\right)|\psi|^2
    \right\} \;.
\end{equation}

Noether's theorem then implies the aforementioned mass conservation:
\begin{equation}
    \label{eq:mass_conservation}
    M(t)
    = \int_0^L \text{d}x \ |\psi(x, t)|^2 
    = \const
\end{equation}
due to invariance of the action under global phase changes $\psi(x,t) \to
e^{i\phi}\psi(x,t)$.

Spatial translation invariance, $\psi(x,t) \to \psi(x-x',t)$, as a consequence
of periodicity and symmetry of the kernel \eqref{eq:kernel} assures conservation
of linear momentum \eqref{eq:kernel} \cite{timmaster}. The total momentum is defined by
\begin{equation}
    P(t) = -\text{Im}\left\{ \int_0^L  dx \ \psi(x,t) \ \partial_x \psi^*(x,t) 
  \right\} = \const
    \label{eq:momentum_conservation}
\end{equation}
for both static ($a(t) = \const)$ and
dynamic background cosmologies ($a(t) \neq \const)$.

Finally, the total energy:
\begin{equation}
    E(a) 
    = \int_0^L \text{d}x \mathcal H = K + W(a),
      \label{eq:total_energy}
\end{equation}  
with
 \begin{equation}
    K = \frac{1}{2}\int_0^L dx |\partial_x \psi(x,t)|^2 \,\, \textrm{and} \,\,
    W(a) = \frac{a}{2} \int_0^L dx \left(\int_0^L dx' G(x,x')|\psi(x',t)|^2\right) |\psi(x,t)|^2
    \label{eq:total_energy1}
\end{equation}
$E(a)$  is only conserved for $a=\const$. It is only then that $\psi(x,t) \to
\psi(x,t-t')$ constitutes a symmetry of Eq. \eqref{eq:action}. 
If background expansion is allowed, the action producing Eq. \eqref{eq:SP1} loses its time 
translation symmetry and Eq. \eqref{eq:total_energy} becomes a function of time. 
The total time derivate is then given by:
\begin{equation}
    \frac{\text{d}E}{\text{d}t} = \frac{\dot a}{a} W(a)
    \label{eq:not_li} \;.
\end{equation}
Note that this result is different from the three dimensional considerations of 
\cite{Kopp2017}, which recovers the well-known Layzer-Irvine equation, 
$\frac{\text{d}E}{\text{d}t} = -\frac{\dot a}{a}(2K+W)$, instead of Eq.
\eqref{eq:not_li}. This is a consequence of the form of the interaction kernel
\eqref{eq:kernel}.

Nevertheless, it is still possible to define a conserved energy, $E_\text{tot}(a)$, by
compensating for the loss implied by Eq. \eqref{eq:not_li}:
\begin{equation}
    \label{eq:conserved_energy}
    E_\text{tot}(a) = E(a) - \int_{a_0}^{a} \text{d}a'\frac{W(a')}{a'} = E(a_0) =
    \const
\end{equation}

\section{Numerical methods}
\label{sec:numerical_method}

We proceed by introducing two independent numerical methods capable of integrating the
nonlinear and nonlocal Schr\"odinger-Poisson equation in an expanding
background cosmology: The pseudospectral Fourier method --- the go-to method of
integrating Eq. \eqref{eq:SP1} in astrophysical settings \cite{Mocz2017,
May2021, May2022} --- and a Crank--Nicolson-based B-spline integrator, popular
for applications in atomic and molecular physics \cite{Bachau_2001}.

Although the temporal integration proceeds differently in both schemes, their
discrete time evolution operators share a set of common features. Most
importantly, both are formally second order accurate in time and \edit{manifestly unitary \cite{NR}}. 
Thus, the suitability of the spatial discretization for Eq. \eqref{eq:SP1} will
determine which out of the two numerical schemes is more appropriate for
simulating Schr\"odinger-Poisson.

As discussed in Sec.~\ref{sec:intro}, the uniform spatial grid used
by the pseudospectral approach is its major bottleneck in large-scale
cosmological simulations, in any number of dimensions. It is for this
reason why we choose to benchmark this "state-of-the-art" approach
against a more adaptive B-spline discretization.

Understanding how this change in the spatial discretization of Eq.
\eqref{eq:SP1} impacts numerical accuracy and computational efficiency will be
helpful to decide, which approach should be developed further, 
in particular to extend our studies in a bottom-up fashion to two spatial dimensions, 
i.e. the $(2+1)$ problem \cite{Navarrete2017}.

\subsection{Pseudospectral or Fourier Method}
\label{sec:PS}

The pseduospectral approach discussed below has emerged, despite high
computational cost for high-resolution simulations, see e.g. \cite{May2021}, 
as the canonical approach for evolving Eq. \eqref{eq:SP1}-\eqref{eq:SP2}. 
The scheme is simple to implement, second-order accurate in time, can be spectrally
accurate in space, is \edit{by construction unitary \cite{NR}}, and approximately energy conserving for
$a(t)=\const$. In case of cosmological applications, it may also be understood
as the wave-dynamic equivalent of the symplectic kick-drift-kick integrator,
omnipresent in $N$-body computations and thus easily integrated into existing
cosmological codes. The following two sections summarize the method's key ingredients.

\subsubsection{Time Integrator}
\label{sec:TD}

The task at hand is to construct a numerically tractable approximation to the
time evolution operator of Eq. \eqref{eq:SP1}. To this end, we follow the
operator splitting technique, i.e. we split Eq. \eqref{eq:SP1} into subproblems
and find ideally exact time evolution operators for the subproblems. A suitable
composition of the latter then yields a time integrator for Eq. \eqref{eq:SP1}.

We realize that Eq. \eqref{eq:SP1} may be split into the subproblems:
\be
i\partial_t \psi = \hat{H}_K \psi \;,
\label{eq:Kterm}
\ee
and 
\be 
i\partial_t \psi = \hat{H}_V \psi.
\label{eq:Vterm}
\ee
While Eq. \eqref{eq:Kterm} and the Poisson equation can be exactly solved in Fourier 
space, Eq. \eqref{eq:Vterm} is easier treated in real $x$ space.

A permissible integrator composition is given by the three-stage Strang
splitting \cite{Strang1968} already used in Ref. \cite{PRD2021}, which explored the fully dynamic case
$a(t)$ mostly by extrapolation of numerical results of the static scenario or
only limited investigation for $a \neq \const$
In this paper, we make progress in treating the expanding case with explicitly
time-dependent $a(t)$ more thoroughly, a much harder scenario for obvious reasons.

The time evolution of the full problem then reads:
\begin{equation}
    \hat{U}_{K+V}(t_0, \Delta t) 
    = 
    \hat{U}_{K}\left(\frac{ \Delta t}{2}\right) \circ 
    \hat{U}_V\left(t_0+\frac{\Delta t}{2}, \Delta t\right) \circ
    \hat{U}_{K}\left(\frac{\Delta t}{2}\right)
    + \mathcal{O}\left(\Delta t^3\right), 
    \label{eq:strang}
\end{equation}
where $\hat{U}_{K}$ and $\hat{U}_{V}$ are the 
time evolution operators of the kinetic and potential sub-problem, respectively.
Since $\hat{U}_K$ can be computed exactly, cf. \eqref{eq:kinetic_problem}, 
it is permissible to combine consecutive kinetic operator applications across
adjacent time steps:
\be
\hat{U}_K(\Delta t/2)\circ\hat{U}_K(\Delta t/2) =\hat{U}_K(\Delta t) \;.
\ee
This reduces the effective number of operator applications.

\subsubsection{Spatial discretization and Time Evolution Operators}
\label{sec:SD}

With the overall structure of $\hat{U}_{K+V}$ in place, we
proceed to specify $\hat{U}_K$ and $\hat{U}_V$.

We choose an equidistant grid of $N$ points in $x$-space, i.e. $x_n = n\Delta x$, and 
the conjugate $k$-space $k_n = \frac{2\pi n}{L}$, $n\in \{0,...,N-1\}$, respectively.
The subproblems are solved separately and $\mathcal{F}$ and $\mathcal{F}^{-1}$ stand 
in the following for the Fourier transform and its inverse.
In Fourier space the kinetic part is solved exactly:
\begin{align*}
    % i\partial_t\psi(x_n,t) &= -\frac{1}{2}\partial^2_x
    % \psi(x_n,t) = 
    %     \mathcal{F}^{-1} diag\left[\frac{k_n^2}{2}\right] 
    %     \mathcal{F} \psi(x_n,t) \\
    \hat{U}_K(\Delta t) &= \mathcal{F}^{-1}
    \mathrm{Diag}\left[\exp{\left\{-i\frac{k_n^2}{2} \Delta t\right\}} \right] 
    \mathcal{F}
%    \psi(x,\Delta t) &\approx \mathcal{F}^{-1} \exp{(\frac{-k^2}{2}
%        \Delta t)} \mathcal{F} \psi(x, 0) .
    \numberthis 
    \label{eq:kinetic_problem}
\end{align*}
The isolated potential part: 
\begin{equation}
    i\partial_t \psi(x_n,t) = 
    a(t)V(|\psi(x_n,t)|^2)\psi(x_n,t)
    \label{eq:potential_problem}
\end{equation}
conserves the norm of $\psi$ component-wise due to the real nature of $V$:
\begin{equation}
    \begin{split}
    \frac{\text d}{\text{d} t} |\psi(x_n,t)|^2 
    &= 
    \psi^*(x_n,t)\partial_t \psi(x_n,t) +
    \psi(x_n,t)\partial_t\psi^*(x_n,t) \\
    &= -i a(t) V(|\psi(x_n,t)|^2) |\psi(x_n,t)|^2 
    + i a(t) V(|\psi(x_n,t)|^2) |\psi(x_n,t)|^2 
    =0 \;,
    \end{split}
\end{equation} 
implying $|\psi(x_n,t)|^2 = |\psi(x_n,t_0)|^2$.
We conclude that the evolution operator of Eq. \eqref{eq:potential_problem},
$\hat{U}_V$, depends explicitly on time only through $a(t)$, and we approximate this 
remaining time dependence by the mid-point method: 
\begin{equation}
    \hat{U}_V(t_0+\frac{\Delta t}{2}, \Delta t) 
    = \mathrm{Diag} \left[
    \exp{ \left\{-i a(t_0+\frac{\Delta t}{2}) 
    V(|\psi(x_n,t_0)|^2 \Delta t \right\} } 
    \right].
    \label{eq:midpoint}
\end{equation}
We apply the convolution theorem and solve the Poisson equation in $k$-space using 
the convolution kernel $-\frac{1}{k_n^2}$, see Eq. \eqref{eq:kernel},
\begin{equation}
    V(|\psi(x_n,t)|^2) = \mathcal{F}^{-1} \mathrm{Diag}
    \left[-\frac{1}{k_n^2}\right]
    \mathcal{F} |\psi(x_n,t)|^2
    \label{eq:convolution_potential}
\end{equation}

We close with two remarks:

Firstly, given that the trivial kinetic subproblem is solved in $k$-space, it
would be desirable to do the same for the potential subproblem as well.
Unfortunately, we cannot efficiently solve the above equations in $k$-space only, 
because of two nonlinearity related problems: (i) in order to calculate the potential $V$, we need the point-wise norm 
$|\psi(x_n,t)|^2$ in $x$-space, not just the wave function itself; and (ii)
$V$ would be highly nonlocal in $k$-space and it is hence
more efficient to apply $\hat{U}_V$ in $x$-space, where it is trivially diagonal.

Secondly, note that (i) the operators in Eq. \eqref{eq:kinetic_problem} and Eq. \eqref{eq:midpoint}
are unitary by construction and (ii) the Strang splitting, Eq. \eqref{eq:strang}, is a \edit{\em composition of single unitary operators} \edit{\cite{NR}}. 
Thus, the overall time integration is guaranteed to be unitary and norm preserving \edit{up to accumulated numerical noise}.
The property of being unitary, just as symplectic integrators for classical Hamiltonian
problems \cite{Sympl2005, Non2023}, allows for a long time evolution with minimal norm loss, which in particular is stable also for
relevantly large time steps as compared to non-unitary integrators. The stability in relatively large windows
of time steps is well known for other nonlinear problems \cite{Mannella1998, PRL2005}, and forms the basis for our method which
has to deal additionally with a fast varying scale factor entering the nonlinear long-range potential.

\subsection{B-spline method}
\label{sec:splines}

In atomic and molecular physics the use of B-spline basis functions is very common 
\cite{Bachau_2001, Bspline-1, Bspline-3, Bspline-4}, since this basis can be often 
adapted to the problem at hand. 
Its advantage is that often the number of basis functions can be reduced to a minimum 
due to the localised nature of the B-splines. 
The number of necessary functions is typically much smaller than the number of grid 
points $N$ in our pseudospectral method. 
In the following we provide a short overview on the method of B-splines and we refer 
to the specialised literature for more details, see e.g.  
\cite{Ratnani, Milanovic, Boor1978}.

\subsubsection{Spatial Discretization}
\label{sec:b-spline-SD}

B-splines are polynomial functions with compact support and cardinal B-splines are 
defined on an equidistant grid. A cardinal B-spline of first order is the function 
\begin{equation}
    \label{eq:spline-1}
    \varphi_{1}(x)= 
    \begin{cases}
        1, & x \in[0,1), \\ 0, & \text { otherwise. }
    \end{cases}
\end{equation}
A cardinal B-spline of order $p, p \in \mathbb{N}$ is defined as the convolution
\begin{equation}
    \label{eq:convolution}
    \varphi_{p}(x)
    =\left(\varphi_{p-1} * \varphi_{1}\right)(x)
    =\int_{\mathbb{R}} \varphi_{p-1}(x-t) \varphi_{1}(t) \mathrm{d} t
    =\int_{0}^{1} \varphi_{p-1}(x-t) \mathrm{d} t,
\end{equation}
or equivalently by the recursion
\be
    \label{eq:spline-2}
    \varphi_{p}(t)
    =\frac{t}{p-1} \varphi_{p-1}(t)+\frac{p-t}{p-1} \varphi_{p-1}(t-1), \quad p \geq 2.
\ee
From Eqs. \eqref{eq:spline-1} and \eqref{eq:convolution} it follows that the Fourier 
transform of $\varphi_p(x)$ is
\begin{equation}
    \label{eq:spline-3}
    \frac{\left(-1+e^{i w}\right)^{p}}{\sqrt{2 \pi} ({\rm i}w)^{p}}.
\end{equation}

\begin{figure}[tb]
    \centering
    \includegraphics[width=0.95\textwidth]{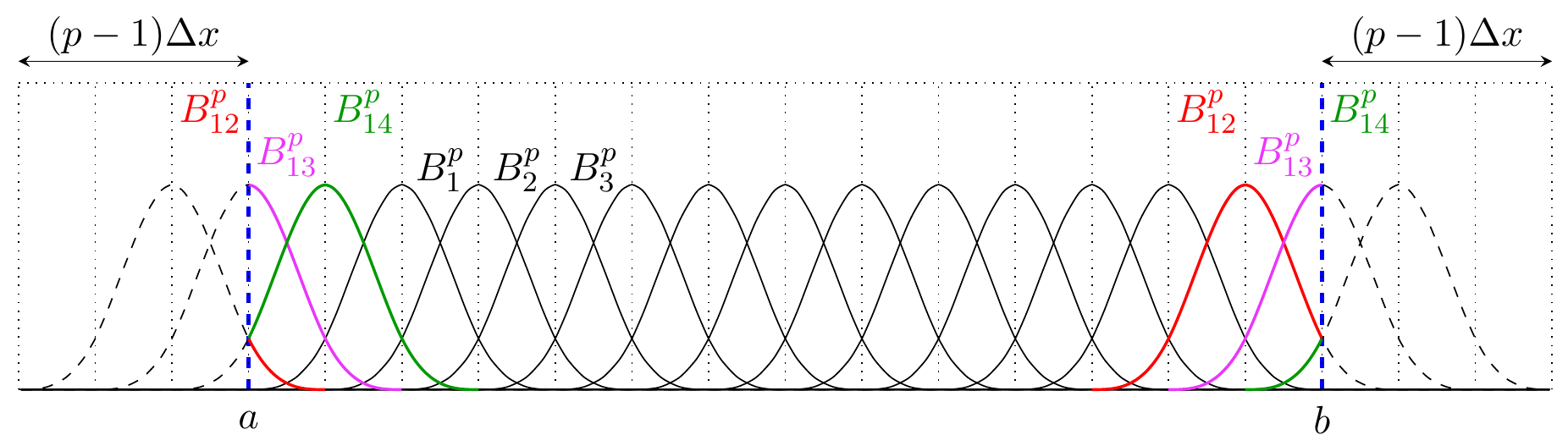}
    \caption{
        A set of $14$ B-splines of order $p=4$ for periodic boundary conditions 
        in the interval $[a,b]$.}
    \label{fig1}
\end{figure}

For a given interval $[a,b]$ and for $j=1,\dots, N$, we define the functions 
\begin{equation}
     \label{eq:spline-4}
    B_j^p(x)= 
    \begin{cases}
        \varphi_p((x-x_j)/\Delta x),& x-x_j\in[a,b],\\
        \varphi_p((x-x_j-b)/\Delta x), & \text { otherwise, }
    \end{cases}
\end{equation}
where $x_j=a+(j-1)\Delta x$, with $\Delta x =(b-a)/N$. 
These functions form a complete basis of $N$ B-splines of order $p$ for periodic 
boundary conditions in the interval $[a,b]$. 
Basic B-spline functions are exemplary shown in Fig. \ref{fig1} for $p=4$ and $N=14$. 
As can be seen in the figure, the B-Spline basis functions are not orthogonal. 
The overlap matrix $\boldsymbol{B}$ is given by the elements
$\langle  B_{j}| B_{k} \rangle= B_{jk} \ne \delta_{jk}$. 
Therefore, the matrix representation  of an operator 
$\hat{A}$ differs from its representation in an orthogonal basis in the following way:
if $\tilde A _{i k}$ are the expansions coefficients of the action of $\hat{A}$ on the B-spline $\ket{B_i}$,
\begin{equation}
\label{eq:spline-5a}
 \hat{A} | B_{i} \rangle  = \sum_k \tilde A _{i k} | B_{k} \rangle
\end{equation}
then
\begin{equation}
\label{eq:spline-5}
 A _{ji}= \langle  B_{j}| \hat{A} | B_{i} \rangle  
 = 
\sum_k \tilde A _{i k} \langle  B_{j} | B_{k} \rangle .
\end{equation}
It is the overlap matrix, which causes off-diagonal elements in $\boldsymbol{A}$. 
More precisely, $\boldsymbol{A}$ consists of a $p+1$-wide band around the diagonal and triangular 
parts of size $p$ in the lower left and upper right corners. 
The main disadvantage of the method is then that the application of a matrix-vector 
product is much more involved compared to a diagonal or a small-banded matrix in an 
orthogonal basis. 
The main advantage of the method is that the number of basis functions typically is 
much smaller than the grid points used in an equidistant spatial basis, such as in 
Sec.~ \ref{sec:PS}. 
The tradeoff between these two aspects will determine the usefulness of the B-spline 
approach. 
Anticipating results from below, we will actually see that for our gravitational 
problem, typically a lot of B-splines are nevertheless necessary such that the method 
becomes very time and memory consuming making it actually less adaptable to our problem.

\subsubsection{Time Integrator}
\label{sec:b-spline-TI}

We evolve the coupled system of equations \eqref{eq:SP1} and \eqref{eq:SP2} using the \edit{norm-preserving}
Crank--Nicolson method \cite{CN1947, NR}. 

\begin{equation} \label{eqn:semicrank}
 \left(\hat{I} +\frac{i}{2}\hat{\overline{\mathrm{H}}}\right)\psi \left(t+\Delta t\right) = \left(\hat{I} -\frac{i}{2}\hat{\overline{\mathrm{H}}}\right)\psi \left(t\right)\,,
\end{equation}
where 
\begin{equation} 
 \hat{\overline{\mathrm{H}}}= \frac{1}{\Delta t} \int^{t+\Delta t}_{t} \hat{H}  dt=  \hat{H}_K +\hat{\overline{H}}_V\,.
\end{equation}
The time-dependent integral $\hat{H}_V$ is calculated with the help of the trapezoidal rule,
\begin{equation} 
\hat{\overline{H}}_V= \frac{1}{2}\left(a(t) V(t) + a(t+\Delta t) V(t+\Delta t) \right)
+ O({\Delta t}^{2})\,.
\end{equation}
Here, the potential $V(t+\Delta t)$ is needed at each time step. 
However, it depends on the unknown wave function $\psi(t+\Delta t)$. 
To circumvent this problem we use a predictor-corrector step, as applied in other 
nonlinear problems \rev{\cite{Mannella1998, PRL2005}}. 
That is, we predict an intermediate solution $\tilde{\psi}(t+\Delta t)$ by applying 
the above mentioned method (\ref{eqn:semicrank}) with 
$\hat{\overline{H}}_V \approx \hat{H}_V=a(t) V(t)$
\begin{equation} \label{eqn:pred}
    \psi (t) \qquad \underrightarrow{\hat{\overline{H}}_V = a(t) V(t)}
     \qquad   \tilde{\psi}(t+\Delta t)\,,
\end{equation}
then we use this predicted wave function to solve the Poisson equation and obtain 
$\tilde{V}(t+\Delta t)$ which we use in the corrector step,
\begin{equation} 
   \label{eqn:corr}
   \psi (t) \qquad \underrightarrow{\hat{\overline{H}}_V
   =\frac{1}{2}( a(t) V(t)+a(t+\Delta t)\tilde{V}(t+\Delta t))} \qquad
   \psi(t+\Delta t)\,.
\end{equation}
In this way we need to solve a system of four uncoupled linear equations at each time 
step \rev{(two each for the predictor and the corrector step, respectively). One corrector step is typically sufficient 
to deal with the time-dependent potential. Similar to the case of the Gross-Pitaevskii equation 
\cite{Mannella1998, PRL2005}, more corrector steps don't significantly improve the outcome, as we checked.}
The Hamiltonian is represented in the non-orthogonal B-spline basis and 
Eq. \eqref{eqn:semicrank} takes the form 
\begin{equation} 
    \label{eqn:crannikol}
    \left(
        \boldsymbol{B} +\frac{i}{2}\overline{\boldsymbol{\mathrm{H}}}
    \right)\psi 
    \left(
        t+\Delta t
    \right) 
    = 
    \left(
        \boldsymbol{B} -\frac{i}{2}\overline{\boldsymbol{\mathrm{H}}}
    \right)\psi 
    \left(t\right)\,.
\end{equation}
This is the most time consuming step in our integrator, since the matrices involved 
in Eq. \eqref{eqn:crannikol} have a banded structure combined with nonzero triangular 
parts in the lower left and upper right corner due to the periodic boundary
conditions. Summarizing, Eq. \eqref{eqn:crannikol}  provides us the time
evolution of our state where the $\overline{\mathbf{H}}$ is calculated using the predictor corrector steps 
\eqref{eqn:pred} and \eqref{eqn:corr}, where the potential $V(t)$ is obtained 
by solving the Poisson equation using a direct matrix inversion of the Laplace 
operator.  As in Eq. \eqref{eq:midpoint}, we approximate $a(t)$ with the mid-point rule. 
 
Altogether, \edit{due to the Crank--Nicolson scheme} also the B-spline method is unitary and hence norm conserving just as the 
pseudospectral method from Sec.~\ref{sec:TD}. Norm (or mass) conservation is crucial in our approach in order to maintain numerical 
errors small and to allow for not too small values of the time step $\Delta t$. 

%%%%%%%%%%%%%%%%%%%%%%%%%%%%%%%%%%%%%%%%%%%%%%%%%%%%%%%%
\section{Numerical Results}
\label{sec:res}

In the following we provide an in-depth numerical comparison of the
pseudospectral method of Sec. \ref{sec:PS} and the B-spline approach,
Sec. \ref{sec:splines}, applied to the numerically challenging problem of integrating the
Schr\"odinger-Poisson equation in an expanding, flat FLRW universe, i.e when
$a(t)$ is given as the solution to the background Eq. \eqref{eq:cosmology}.  
Sec. \ref{sub:res-ps} establishes PS as our
baseline scheme by a comprehensive study of its convergence and accuracy as a function of the 
spatio-temporal grid. The obtained high-resolution reference solutions are then
used in Sec. \ref{sub:comparison} for an exhaustive comparison against the
independently implemented Crank--Nicolson B-spline approach: Sec.
\ref{subsub:power} investigates deviations and systematics in the nonlinear matter power
spectrum, Sec. \ref{subsub:mass} - \ref{subsub:energy} assess the adherence of
both schemes to the conservation properties introduced in Sec.
\ref{sub:conservation_laws}. We close by benchmarking both schemes in terms of CPU 
runtime and memory in Sec. \ref{subsub:performance}.

For reference, the chosen box sizes are in our dimensionless units $L=100, 500,$ and $1000$, 
corresponding to cosmologically relevant sizes of $L\approx 2, 10,$ and $20\,\si{\Mpc}$ 
for the parameters reported in Sec.~\ref{sub:SPE}. The typical distance between galaxies is of $\mathcal{O}(\SI{1}{\Mpc})$, for instance.

For small $L$, the initial state, as described in Sec. \ref{sub:ic}, is mostly a single period 
sinusoidal function. The amplitude fluctuations for these states firstly rises to 
$\delta\sim0.15$ at $L\approx100$. 
Beyond that, the number of periods grows to make the state highly oscillating. 
All simulations depart from an initial coupling strength of $a(t_0)\approx0.01$ at initial time $t_0=0$.

\subsection{\label{sub:res-ps}Pseudospectral Method}

\begin{figure}[tb]
    \centering
    \includegraphics{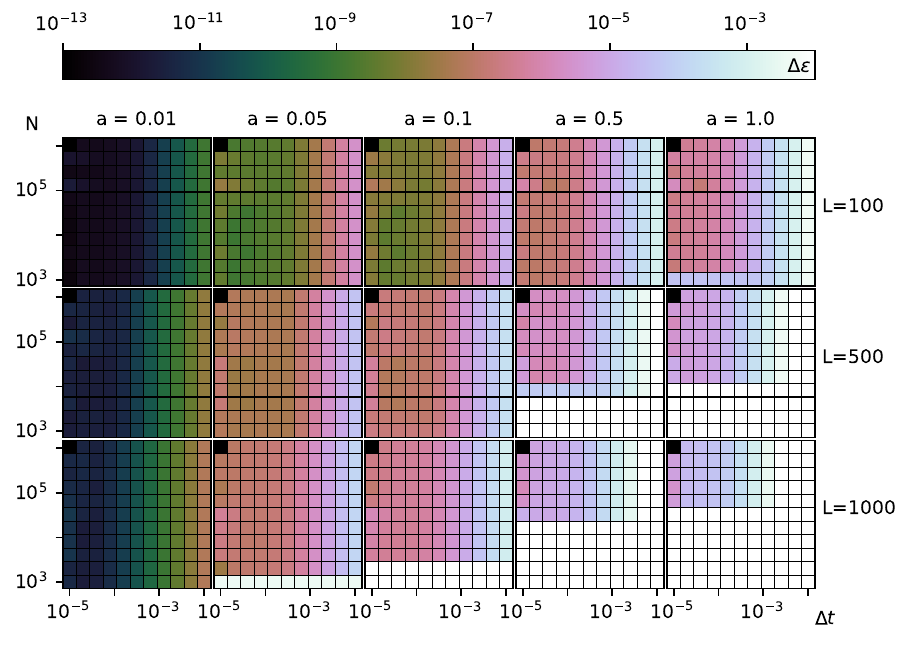}
    \caption{ 
    Convergence of the PS method with respect to different parameters in the (1+1)-SP.
    Each cell colour-maps the errors $\Delta\epsilon$ with respect to the 
    black-square reference.
    The three box sizes $L\in\{100, 500, 1000\}$ are plotted row-wise.
    The time development is given column-wise in terms of the scale factor $a$.
    The $N$-grid is in powers of 2 to ensure overlapping spatial nodes.
    To make it easier to read, $N$ and $\Delta t$ have ticks at powers of 10.
    Errors induced by constraints in $\Delta x$ show a sharp edge in $N$-direction, 
    which moves toward higher $N$ as the nonlinearity increases.
    Errors in the time domain on the contrary create a smooth saturation with 
    decreasing $\Delta t$ that does not move with time.
    }
    \label{fig:2}
\end{figure}

The first point that we want to check here is the convergence of the PS method
with respect to the systems size in the number of points in space and time. 
Hence we look for the convergence plateau in both grid spacings $(\Delta x, \Delta t)$.
A fine-grained solution for the parameters $N=2^{20}$, $\Delta t=10^{-5}$ serves as 
reference relative to which errors are calculated. 
Figure~\ref{fig:2} shows the error $\Delta\epsilon$, which is computed by taking the 
relative Euclidean norm:
\begin{equation}
    \Delta\epsilon(t) = \frac{\left(\int_{0}^{L} 
    |\Psi(t) - \Psi_{\text{ref}}(t)|^2 \,dx \right)^{\frac{1}{2}}}
    {\left(\int_{0}^{L} |\Psi_{\text{ref}}(t)|^2 \,dx\right)^\frac{1}{2}}.
    \label{eq:L2}
\end{equation}
The black-squares indicate our reference solution in Fig.~\ref{fig:2}. 
The parameter space spans three orders of magnitude in
the space domain $N\in\{2^{10},..,2^{20}\}$ as well as in the time domain 
$\Delta t \in \{2^0,...,2^{10}\}\cdot10^{-5}$.
Three box sizes are plotted row-wise. $a(t)$ is the natural choice to indicate the 
time evolution since it couples the nonlinear potential to 
the Schr\"odinger equation and therefore provides the best physical insight. 
For example, we can expect matter fluctuations to be dispersed randomly 
at $a\ll1$ and to clump at $a\to1$.

Errors induced by the space grid and time grid are clearly distinguishable in Fig.~\ref{fig:2}. 
On one hand, an insufficient Fourier basis manifests in a sharp edge of $\Delta\epsilon$ in 
the $N$-direction ($y$ axis). 
The problem becomes more nonlinear either when going to a larger box size or with the increasing
scale factor $a(t)$ during the integration. This implies increasingly compact density
distributions in $x$-space and hence a wider distribution in $k$-space, which has to be 
compatible with the maximal resolved wavenumber set via $L/N$.
The minimum requirement $N_{min}$ for a converged solution is such that all modes in the power 
spectrum are resolved. That is the case where the natural roll-off dives into the noise floor 
of numerical precision. We refer here already to Sec.~\ref{subsub:power}, in which we
discuss in more detail the intricacies of the spatial representation, see in particular 
Fig.~\ref{fig:5} there.

\begin{figure}[tb]
    \centering
    \includegraphics{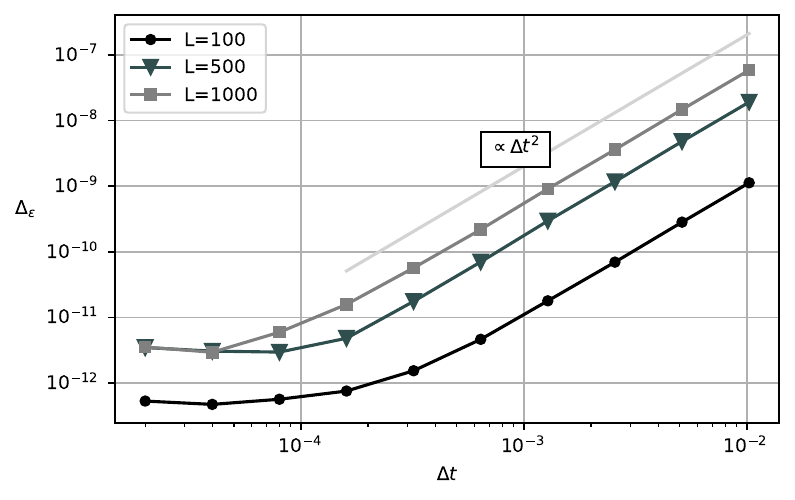}
    \caption{Scaling of the convergence of the Fourier--Strang integrator as a function of time step $\Delta t$ for the box sizes shown in the legend. As expected from the 
    second-order scheme, the plateau is approached as $\Delta_\epsilon \propto \Delta t^2$.}
    \label{fig:3}
\end{figure}

Although a smaller grid $N < N_{min}$ does not resolve all of the small-scale dynamics, 
especially as $a\to1$, the integrator still stays stable. On the other hand, 
$\Delta\epsilon$ approaches the convergence plateau in $\Delta t$ in a quadratic manner 
$\Delta_\epsilon\propto\Delta t^2$, as the second-order splitting scheme suggests, 
see Fig.~\ref{fig:3}.
Even though the absolute value of the errors change, the shape of temporal convergence stays 
approximately constant.
We deem $\Delta\tau=8\times10^{-5}$ the first temporal grid point distance inside the 
convergence plateau.

One has to consider that if the box size is changed, the different degree of 
nonlinearity drastically effects the evolution dynamics.
As the box size increases, the total mass increases due to the normalization condition, 
see Eq.~\eqref{eq:norm}, and the local potential wells become deeper on average, 
which is a prerequisite for more dynamics of the wave function. 
At the same time, this adds to the computational cost, because more Fourier modes are needed to 
resolve the appearing peculiar local velocities  $v_{\rm pec}$, 
which are defined by local phase changes of the wave function. 
In Madelung representation we have 
\begin{equation}
    v_{\rm pec} 
    = \partial_x S(x,t), \quad \text{with}  \quad \psi (x,t) 
    = \sqrt{\rho (x,t)}\exp(i S(x,t)).
\end{equation}
Since the exponential function is $2\pi$-periodic, we need to restrict phase jumps between 
adjacent grid points to $\partial_x S < 2\pi/\Delta x$, which in fact is the maximal 
phase velocity we can numerically faithfully simulate.

\subsection{Comparison of the pseudospectral and the B-spline method}
\label{sub:comparison}

With the scope of high computational performance in mind, two aspects can be checked for 
improvements, the spatial and the temporal representation of the problem. 
Regarding the former, we compare our PS method now with the B-spline method introduced in 
Sec.~\ref{sec:splines}.
Our B-Spline method has been implemented independently by the Cal{\`i} group which makes the 
comparison with the PS technique developed in Parma/Heidelberg very valid. 

\begin{figure}[bt]
    \centering
    \includegraphics{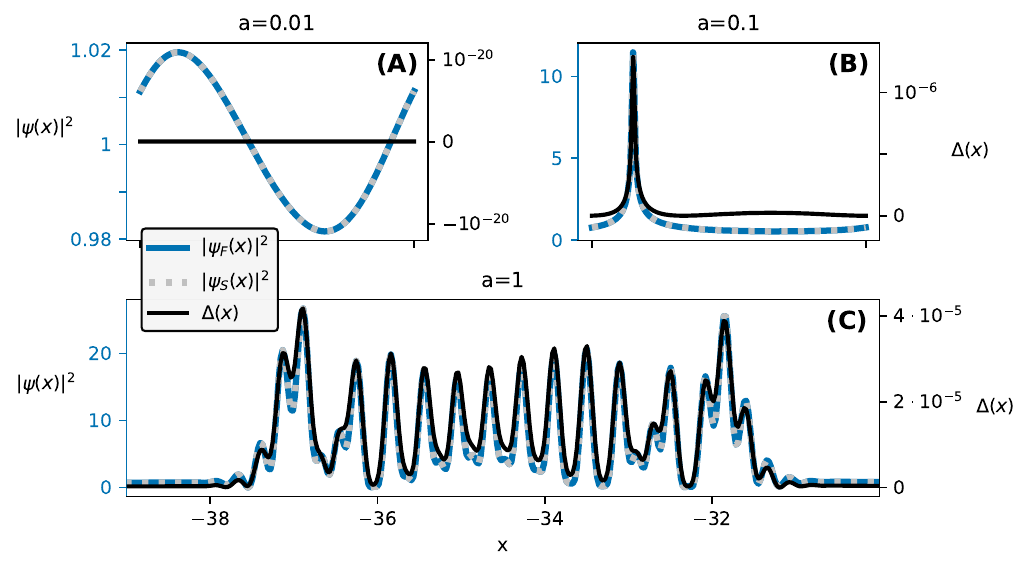}
    \caption{
   Wave function in a small box (L=100)
    of the B-spline (grey dashed line)  as well as the Fourier (blue solid line) method and 
    their difference $\Delta$ (black solid line) at various stages of 
    the integration given in terms of the scale factor 
    $a=0.01$ (A), $a=0.1$ (B), and $a=1$ (C).
     $\Delta$ is limited by 
    numerical precision in the beginning of the 
    simulation and grows to $\mathcal{O}(10^{-5})$ at the end at $a=1$ (C).
    We further see that after some short initial stage the differences in the two methods are approximately 
    proportional to the density itself (B, C), as explained in the main text. 
    }
    \label{fig:4}
\end{figure}

At first measure, we directly compare the wave functions from the PS $\psi_{\rm PS}$ and the 
B-spline method $\psi_{\rm BS}$ via their distance:
\begin{equation}
    \Delta(x, t) = |\psi_{\rm PS}(x, t) - \psi_{\rm BS}(x, t)|^2 \;.
\end{equation}
In Fig.~\ref{fig:4} both wave functions are shown in thick blue, as well as their 
difference at various stages of the integration by the dashed lines, given in terms of the 
scale factor $a(t)$. Please note the difference in scale between the two.

At the simulation start, at $a\approx0.01$ corresponding to a redshift $z=100$, numerical errors of 
$\mathcal{O}(10^{-20})$ or below appear, which are within our numerical precision 
(standard double precision was used throughout). 
At $a=0.1$, the densest region, on the left of the panels in Fig.~\ref{fig:4}, begins to 
collapse due to the gravitational attraction. 
And as expected, in the areas of depletion and accumulation of mass, the differences between 
the codes begin to show. 
Both solutions are always close and the difference grows to maximally $\mathcal{O}(10^{-5})$ at 
the end of the simulation, i.e. at $a=1$. 
We can say that the two independently developed methods give a reasonably good agreement. 
This is so far the best confirmation of the quality of our methods, next to self-consistent 
checks, such as the ones presented in the next sections.

\subsubsection{Power Spectra}
\label{subsub:power}

\begin{figure}[tb]
    \centering
    \includegraphics[width=\linewidth]{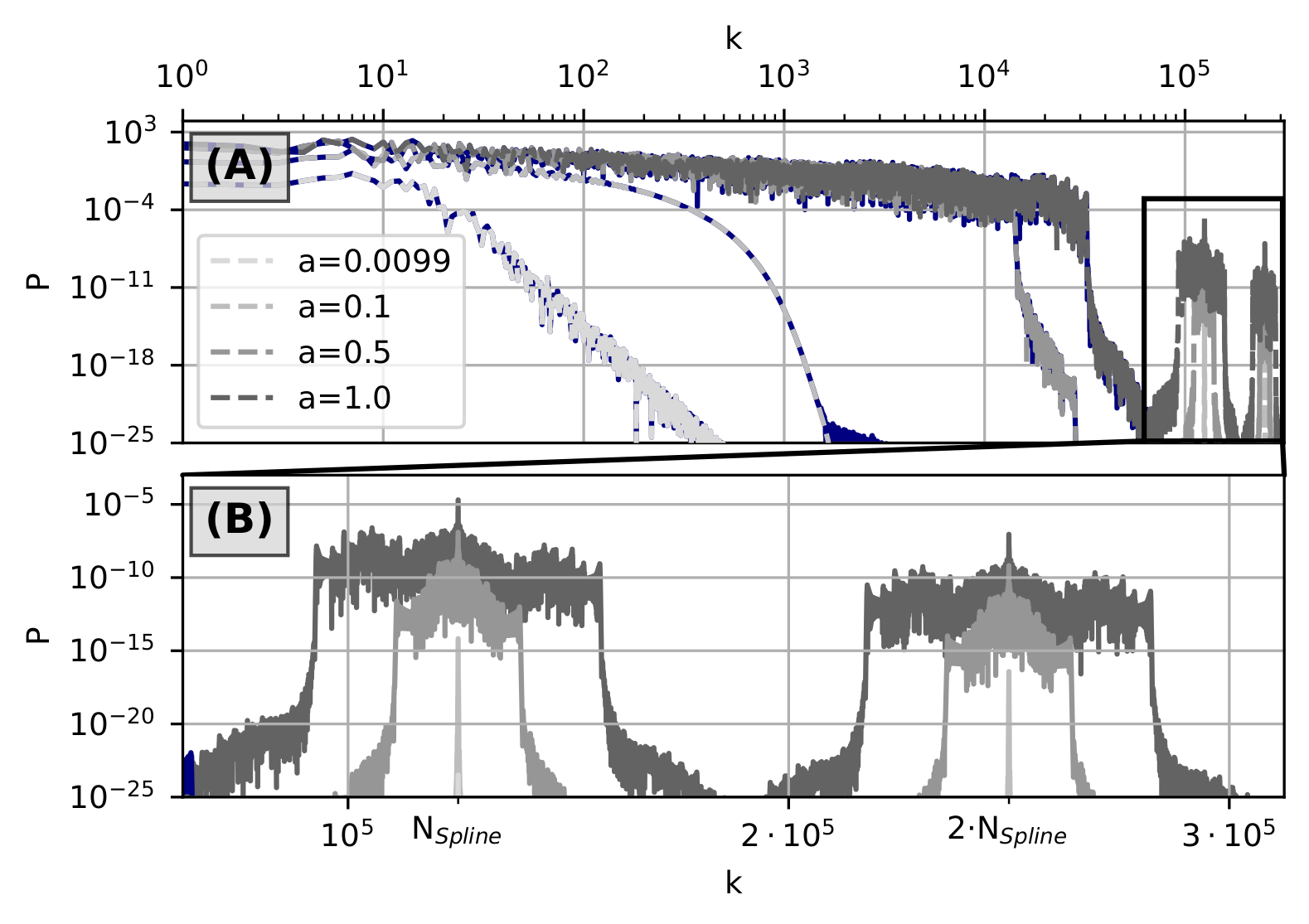}
    \caption{(a): Evolution of the power spectra 
    in the largest box that we study here, with $L=1000$, in terms of the scale factor $a(t)$, see legend.
    The Fourier (PS) method (blue) underlies the B-splines in grey color. 
    Modes $k<N_{\rm Spline}/2$ coincide in their spectra.
    (b): Zoom into the 'oversampled' B-Splines.
    Artificial peaks show up at multiples of the grid size at $k=nN_{\rm spline}$,
    $n\in\mathbb{N}$ (black dashed lines).
    They exist early on and expand farther during integration.} 
    \label{fig:5}
\end{figure}

Going to larger box sizes $L\in[500, 1000]$ shows the weakness of the B-spline method. 
This is best seen when looking at the power spectra of the density fluctuations. 
Figure \ref{fig:5} presents the power spectra for positive modes only since the spectra are 
symmetric around zero for real densities. 
The PS results are shown in blue underlying the B-spline ones in grey. 
For this plot, the number of B-spline basis functions $N_{Spline}=125000$ was very large, 
in such a way, that a Fourier basis of that size can resolve the entire spectrum.

\begin{figure}[tb]
    \centering
    %[width=12cm, height=5cm]
    \includegraphics[width=\textwidth]{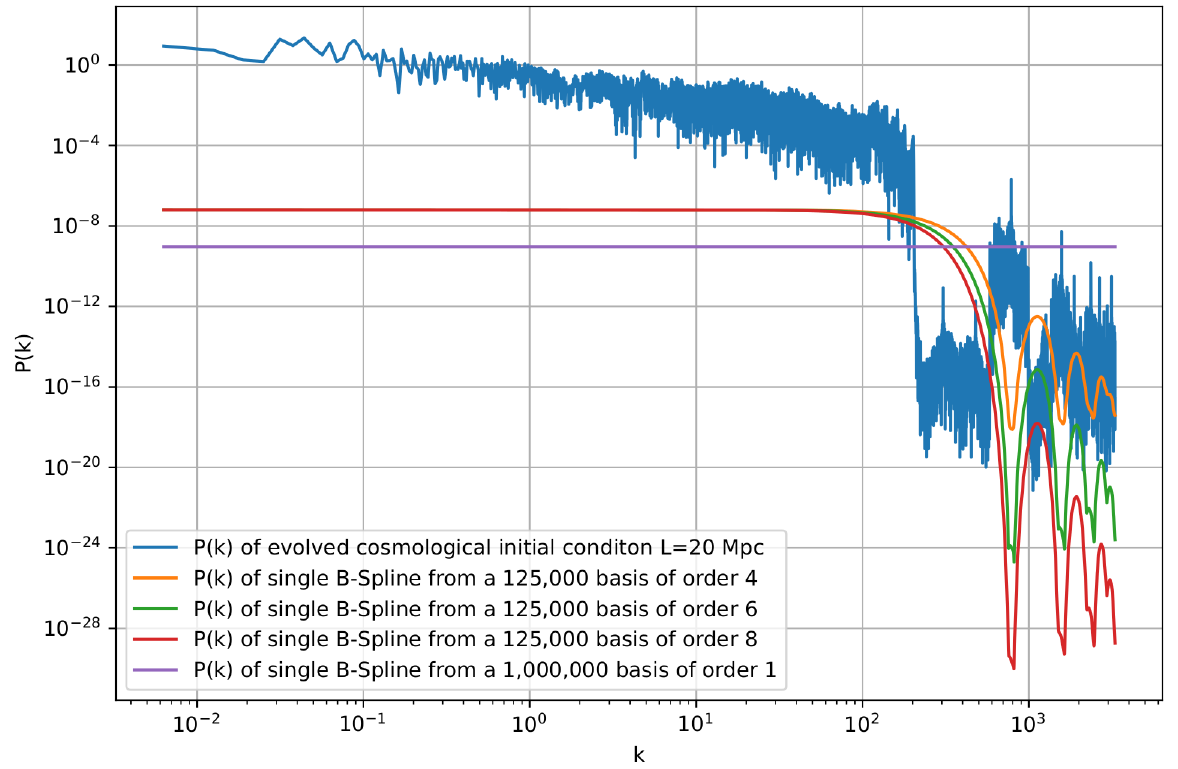}
    \caption{Power spectra of (i) the state in the large box (L=1000)
    at the end of the simulation (blue line), (ii) a single B-spline
    on a $N=10^6$ grid with order 1 (purple line), and (iii) single B-splines
    on a $N=125000$ grid with order 4 (orange line), 6 (green), and 8 (red).
    At multiples of the grid size $N$, the single
    splines have minima in their power spectrum,
    e.g. at $125k$ for the green curve.
    That is the place, where power is generated
    in simulation runs (blue curve).
    Splines of different order exhibit their minima at
    the same position.
    Whether the additionally generated power is negligible depends
    on how far outside of the natural roll-off of 
    the power spectrum these minima are,
    e.g. the first minimum of the purple B-spline is well outside the
    depicted range.
    }
    \label{fig:6}
\end{figure}

In the lower part of the spectrum $k < \frac{N_{\rm Spline}}{2}$, both power spectra follow 
each other closely as it should be.
We observe, however, additional peaks in the B-spline spectra at 
$k=nN_{\rm spline}, n\in\mathbb{N}$, which exist early on and grow in time not only in amplitude,
but also in extent. 
They originate in an inherent lack of resolution of the B-Spline basis at those scales, 
as can be seen from the Fourier transform Eq. \eqref{eq:spline-3}, see Fig.~\ref{fig:6}.
To ensure convergence of the result, the spot where the B-splines have a dip in resolution, 
must be far outside the natural roll-off of the power spectrum.
A Fourier (PS) method on the contrary can be limited to the roll-off and seems therefore be 
better suited to solve the Schr\"odinger-Poisson equation than the B-spline approach that must 
include a very large number of basis functions to avoid problems in the momentum-space 
resolution.
This goes against the expectation that a B-spline method would demand only a small basis size. 
As a consequence its main advantage, see Sec.~\ref{sec:splines} is lost, and the method 
works efficiently only for small box sizes. Currently, it is not efficient to use B-spline 
basis sizes greater than $150k$ due to the high CPU time required. 
However the B-spline code still has some room for improvement since its routines 
could, in principle, be parallelized. 
There seems to be only little improvement possible by a further optimisation of the B-Spline 
parameters to enhance the spatial resolution. 
Alternatively, one may try to use a different set of basis functions, or some kind of 
adaptive mesh refinement \cite{Schive2014, Mina2020, Shive2018, Veltmaat2018, enzo, Schwabe2020}.

\subsubsection{Conservation of Mass}
\label{subsub:mass}

A physical way to assess the accuracy of our integrators is by checking how well norm or the 
total mass is conserved.
We can expect good mass conservation, since the used integrators are unitary by design.
We measure the absolute deviation $\Delta M(t)$ as, cf. eq. \eqref{eq:mass_conservation}:
\begin{equation}
    \Delta M(t) = \Delta x \sum_n \left\{ |\psi(x_n, t)|^2 - |\psi(x_n, t_0)|^2  \right\} .
\end{equation}

\begin{figure}
    \centering
    \includegraphics{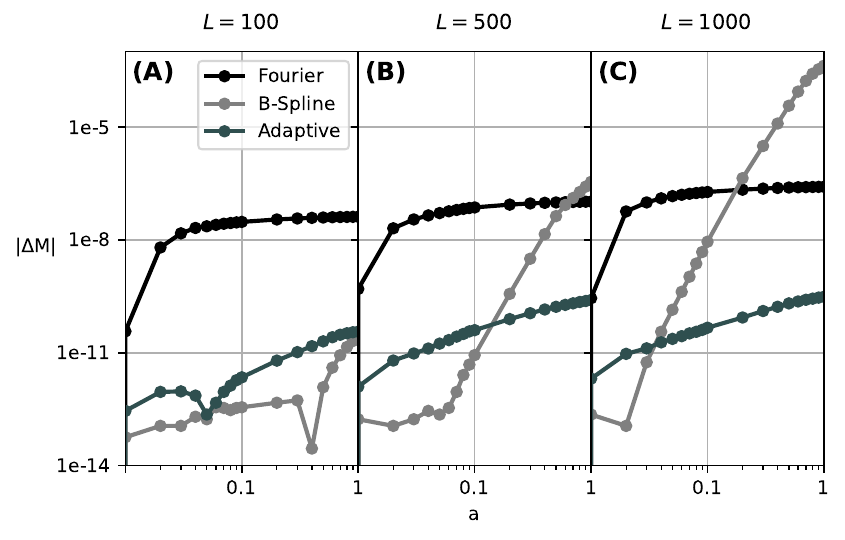}
    \caption{Mass conservation of the PS, the B-Spline and 
    the new adaptive method (to be discussed in Sec. \ref{sec:adaptive_stepper}) 
    for three box sizes.
    The Fourier method conserves the total mass below the $10^{-6}$ range, whereas
    the B-spline method is in advantage for smaller $L$, but struggles with the larger
    boxes, where $|\Delta M|$ grows to 
    $\mathcal{O}(10^{-3})$.
    The adaptive method seems to provide the best solutions. 
    It starts at $|\Delta M| < \mathcal{O}(10^{-12})$ and 
    stays bounded by $|\Delta M| = \mathcal{O}(10^{-9})$.}
    \label{fig:7}
\end{figure}

Figure~\ref{fig:7} shows how the mass evolves with time, 
beginning at $|\Delta M| \approx 0$, where approximately means below $\mathcal{O}(10^{-12})$. 
The upper row derives from the PS method, the lower row from the B-Spline approach. 
The three panels reflect the different box sizes $L=100, 500$, and $1000$.
The PS method conserves the mass well ($|\Delta M| < 10^{-6}$) in all cases, 
while the B-spline method struggles with large box sizes where the error becomes 
$\Delta M = \mathcal{O}(10^{-3})$.
It is worth mentioning the correlation between the increase of mass and the generation of 
power above the noise floor in Fig.~\ref{fig:5}, as discussed in Sec. \ref{subsub:power}.

\subsubsection{Conservation of Momentum}
\label{subsub:momentum}

The discrete version of Eq.~\eqref{eq:momentum_conservation} takes more thought. 
Here the results may be very sensitive to how the derivative is actually computed,
especially when low-order finite differencing schemes are applied. We take the most accurate 
way and use the spectral derivative:
\begin{equation}
    P(t) = \Delta x \sum_n \left\{
    \mathcal{F}^{-1} k_n \mathcal{F}
    \psi^*(x_n, t) \psi(x_n, t) \right\}.
\end{equation}
While for our PS results, the situation and the error is similar to the case of mass 
conversation discussed above, we observe that the B-Spline approach leads to relatively large 
errors in the momentum conservation, in particular for larger box sizes and larger integration 
time, i.e. larger $a(t)$, even going above $10^{-1}$ for $a \to 1$ at $L=500$ and $1000$.
Those relatively large errors arise from taking the spatial derivative in space representation, 
necessary since the B-spline integrator exclusively works in the spatial $x$-representation, 
see Sec. \ref{sec:splines}.
We have checked this by explicitly deriving our SP results in space representation in a similar 
manner, either with a three-point or a five-point method, which for large $a(t)>0.1$ induces a 
similar increase in error.

\begin{figure}[tb]
    \centering
    \includegraphics{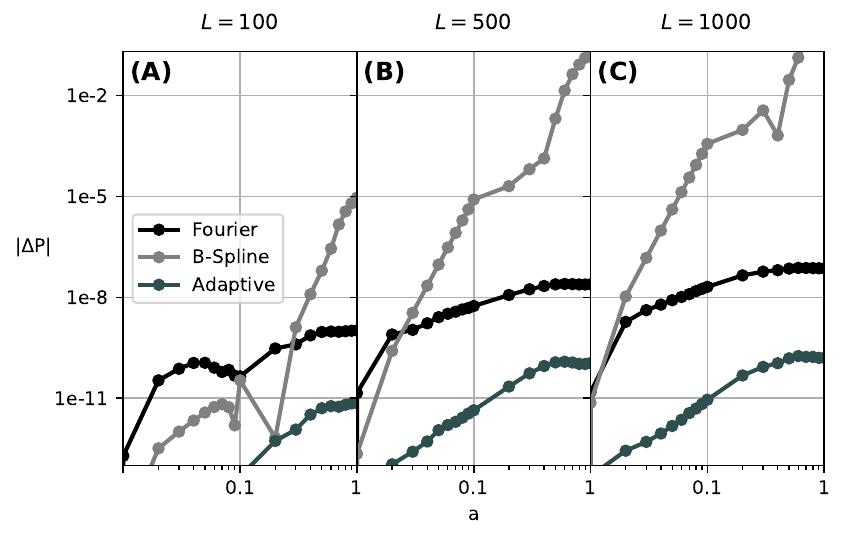}
    \caption{Momentum conservation of the PS, the
    B-Spline and the adaptive method, see Sec. \ref{sec:adaptive_stepper}, for the three 
    box sizes. 
    The situation is qualitatively similar to Fig.~\ref{fig:7}, but problems 
    in the B-Spline method arise earlier than in the mass because the momentum must be
    obtained from the spatial representation of the B-splines.}
    \label{fig:8}
\end{figure}

\subsubsection{Energy Evolution}
\label{subsub:energy}

\begin{figure}[tb]
    \centering
    \includegraphics[width=1.\textwidth]{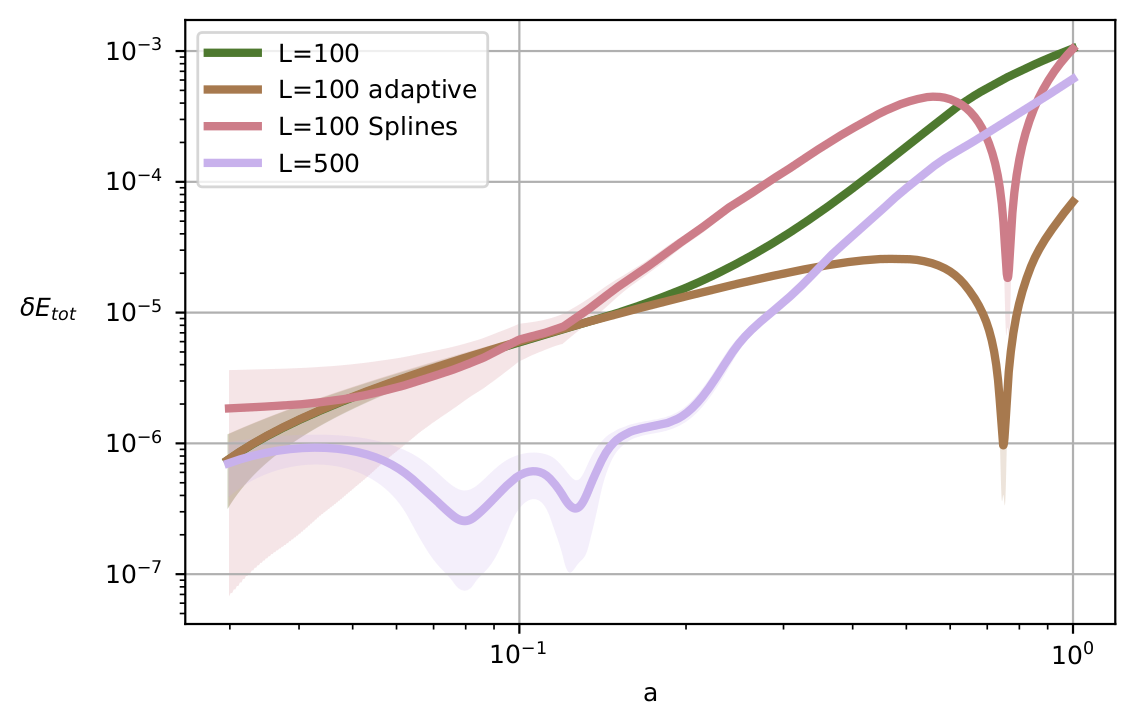}
    \caption{Energy test from Eq. \eqref{eq:new-e-test} vs. $a(t)$ for box sizes $L=100$ and $L=500$. In the former case, results from the PS (dark green and brown lines)
    and the B-spline (wine red line) methods are compared. Local averages over small windows are taken along a uniform grid in $a(t)$ with standard deviation (shadowed areas). As in previous plots, the time-adaptive
    PS (brown line) tends to work best, but all methods show a rather small maximal error of $10^{-3}$ at $a=1$. 
    The minima shortly before in the brown and wine-red curves are due to sign changes
    of the argument in the modulus in Eq. \eqref{eq:new-e-test}.
    }
    \label{fig:9}
\end{figure}

Figure \ref{fig:9} depicts the relative deviation 
\be
\label{eq:new-e-test}
\delta E_\text{tot} =
|E_\text{tot}(a)/E(a_0) - 1|,
\ee 
cf. Eq. \eqref{eq:conserved_energy} as a function of the coupling constant $a$ for
various box sizes and integrators, including the time-adaptive extension of the
PS method, which we present in Sec. \ref{sec:adaptive_stepper}. To arrive at
this result, and in line with the computation of the linear momentum, we compute
the instantaneous kinetic energy by evaluating the spatial derivative in $k$-space for the PS method.
The temporal integration in Eq. \eqref{eq:conserved_energy} is approximated by a more 
physical uniform sampling in $a(t)$, rather than in $t$. All methods show a rather small maximal 
relative deviation of $10^{-3}$, which is only reached at the final integration time, i.e. at $a=1$.

\subsubsection{Numerical performance tests}
\label{subsub:performance}

\begin{figure}[tb]
    \centering
    \includegraphics{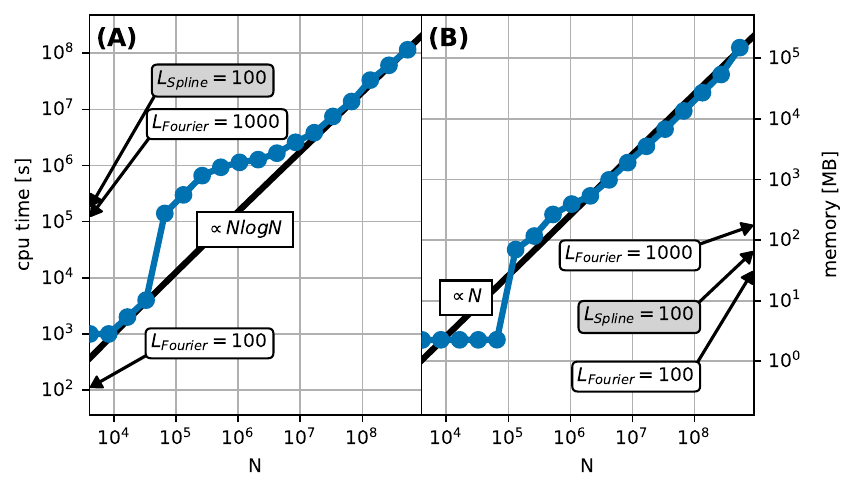}
    \caption{Total CPU time and memory requirements as
    a function of the basis size $N$. 
    The expected scaling $\propto N \log N$ and
    $\propto N$ is recovered for large $N$.
    Small basis simulations are more efficient, as
    long as important vectors fit into the processors
    cache (SRAM).
    Runs with middle sized $N$ scale become worse, because
    the limiting factor is now the memory access.
    The left half shows data, which was mapped
    in a linear fashion to match a time step
    $\Delta t = 10^{-5}$.
    The arrows mark the minimal requirements
    for the specific problem sizes $L=100$ and
    $L=1000$ from our previous figures.}
    \label{fig:10}
\end{figure}

With the cosmological background in mind, the above discussed simulations represent reasonable box sizes of a few 
\si{Mpc}--the typical distance between galaxies--to a few tens of \si{Mpc}. All PS based simulations were executed on the
bwUniCluster thought Heidelberg University\footnote{Hardware specifications are listed at \url{https://wiki.bwhpc.de/e/BwUniCluster_2.0_Hardware_and_Architecture}}
with Intel Xeon Gold 6230 processors with \SI{1.25}{\MB} of L1 cache, that supports basis sizes up to $N\approx 2^{14}$ without additional RAM.
Fourier transforms are sped up with the Fast Fourier Transform (FFT) package FFTW and parallelized with OpenMP on 16 cores.
The computational resources needed for those simulations are shown in Fig.~\ref{fig:10}. 
With a time step $\Delta t = 10^{-2}$, much larger than appropriate from the convergence study, even hypothetical basis sizes of $\mathcal{O}(10^8)$ would be feasible. 
The CPU time can be linearly rescaled if smaller time steps are used, while the memory consumption is unaffected by a change of $\Delta t$ only.
As for FFTs expected, the CPU time approaches a scaling  $\propto N\text{log}N$ and the memory (RAM) a linear scaling $\propto N$ at large basis sizes. 
At medium basis sizes ($N=10^5, 10^6$), unfortunately caching limitations play a significant role in time consumption.

The bwUniCluster's main ('Thin') nodes have a memory capacity of $\mathcal{O}(\SI{100}{\GB})$, which limits us to basis sizes $N\leq10^9$.
CPU time and memory have plenty of headroom for our one-dimensional simulations. The situation would surely change in higher dimensions.
In 3D for example this means $10^3$ grid points in every dimension. That is why many Fourier based approaches only handle 
resolutions up to $N=1024^3$, see also \cite{Mocz2017, Woo2009}. Whether memory consumption will become a limiting factor 
depends on how the physical behaviour changes when we move to higher dimensions.
Since the interaction potential is more confined in 2D and 3D, and the force-law is not constant as in 1D,
we have reason to belief that the situation is less problematic than an upscaled 1D version.
However, \cite{Schive2014} reported a dynamic range of $\mathcal{O}(10^5)$ to be necessary to resolve compact
halo-cores. On the one hand, a box size of a few $\si{\Mpc}$ is needed to accumulate enough mass for the formation of a halo core, i.e. in order to simulate cosmologically interesting evolutions.
On the other hand the spatial resolution must be at least a few tens of $\si{\pc}$ to capture highly oscillating wave functions.

Bigger nodes or larger supercomputers could deliver a maximum performance upgrade of $\mathcal{O}(10-100)$. That might still not be enough for full-fledged three dimensional simulations, which is why we investigate an optimization of performance in the next Sec. We will see there that with an optimized time stepper a speed-up factor of a few tens can be gained in CPU time for our (1+1) SP problem.

\section{Optimized time stepping}
\label{sec:adaptive_stepper}

While working well for small box sizes and for a small number of basis functions, our B-spline integrator, presented in Sec. \ref{sec:splines}, unfortunately turned out to be less performing than our PS or Fourier evolution code, see Sec. \ref{sec:PS}. Hence, in the following we exclusively discuss the PS method.

Besides trying to change the spatial grid representation, we attacked the temporal discretization. The first improvement we had tried was a higher-order integrator with respect to the second-order String splitting reported above. While obviously more expensive in the necessary number of operators, such a higher-order integrator might allow for larger time steps and thus overall result in a gain in the CPU time. Our original code with a fixed time step, as the one reported in Sec. \ref{sec:PS}, was not able to give significant gains. However, adding a possibility for an adaptive time step, we could come up with significant improvements which will be reported in the following.

\subsection{Fourth-order integrator}
\label{sec:4th_order}

A generalized splitting scheme of order $p$ has the following form:
\begin{equation}
    \mathcal{U}_{K+V} = \mathcal{U}_K(b_s\Delta t) \circ 
    \mathcal{U}_V(t_{s-1}, a_{s-1}\Delta t)  \circ ... \circ 
    \mathcal{U}_{V}(t_{1}, a_1\Delta t) \circ \mathcal{U}_{K}(b_1 
    \Delta t)
    + \mathcal{O}(\Delta t^{p+1}),
    \label{eq:splitting}
\end{equation}
where $a_i$ ($i=1,2,\ldots,s-1$) and $b_i$ ($i=1,2,\ldots,s $) are the splitting coefficients and $t_i=t_0+\sum_{j=1}^{i}b_j \Delta t$, 
In comparison, in Eq.~\eqref{eq:strang} the coefficients were $a_1=1, b_1=b_2=\frac{1}{2}$.
$s$ is the number of stages, which not necessarily coincides with the accuracy $p$ (this is true only for certain low-order schemes).
A plethora of higher-order integrators is on the market, see e.g. \cite{auzinger2016practical} that gives some insight
on the construction of these schemes, and we refer to \cite{Auzinger_web} for a collection of splitting coefficients. \edit{Please note that any scheme of the
form as shown in Eq. \eqref{eq:splitting} is again norm preserving since each single evolution over one time step is, independently of the precise value of the step.}

We chose a fourth-order method, since it offers a significant improvement in accuracy while at the same time being compatible with an adaptive time-step
selection explained in Sec. \ref{sec:adaptive}. In particular, we use the splitting scheme "Emb 4/3 BM PRK/A" from \cite{auzinger2016practical}: It is an embedded
scheme, meaning it has two co-working integrators which can be compared along the integration. One is third-order accurate ($s=4$ stages) and acts as the 'controller', the other one is fourth-order accurate ($s=7$ stages) and acts as the 'worker'. Both share the first three splitting coefficients, which reduces the effective number of stages.
Instead of two Fourier transforms per time step in the Strang splitting reported in Sec. \ref{sec:TD}, 17 (sic!) of them are necessary in the new scheme. 
That includes the 'first same as last' (FSAL) property, where consecutive kinetic operators $\hat{U}_K(b_{\rm last}\Delta t) \cdot \hat{U}_K(b_1\Delta t)$
are directly applied without the need of an in-between Fourier transform. 

An adaptive time-step selection is then implemented by comparison of the fourth- and third-order integrator. This comparison gives an intrinsic time-dependent error estimate.
The advantage of such an intrinsic error estimation is that other checks, such as based on the famous Courant-Friedrichs-Levi (CFL) criterion for many finite differencing schemes \cite{NR, CFL1928}, as applied e.g. in \cite{Shive2018, Mina2020, Li2019} in our context, are not required.

\begin{figure}[tb]
    \centering
    \includegraphics{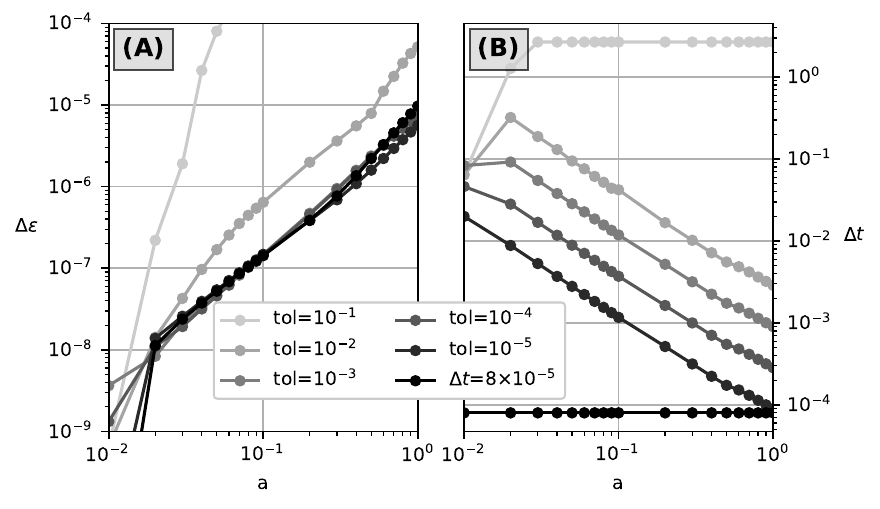}
    \caption{ 
    Left panel (A): Deviation $\Delta \epsilon$, see \eqref{eq:L2}, from the reference solution (Strang splitting, $\Delta t=\text{const.}=10^{-5},  N=2^{20}$)
    as a function of stepping parameter $tol$ (see the legend for its values) of our adaptive time stepper. All data are for the $L=500$ box. 
    Five adaptive, 4th-order solutions with different $tol$ parameters are shown, and one converged 
    second-order solution with a constant time step $\Delta t=8\times10^{-5}$.
    Right panel (B): Evolution of the time step $\Delta t$ during integration for the adaptive stepper. Time-step selection works well for 
    $tol\leq10^{-2}$.}
    \label{fig:11}
\end{figure}

\subsection{Adaptive Time Stepper}
\label{sec:adaptive}

\begin{figure}[tb]
    \centering
    \includegraphics{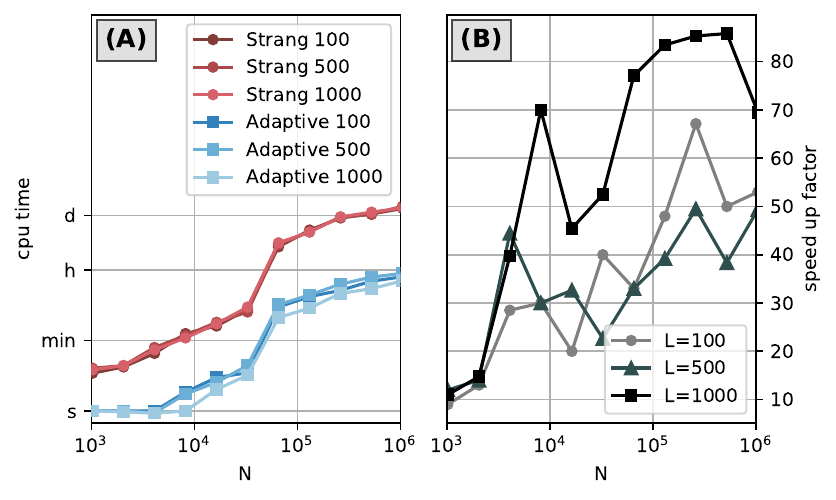}
    \caption{ 
    (A): Required CPU of the three boxes sizes
    $L\in\{100, 500, 1000\}$ (see legends) for the PS method with Strang splitting (red lines; using a fixed $\Delta t = 8\times10^{-5}$) and with adaptive time stepper (blue lines).
    The tolerance parameters were chosen as $tol\in\{10^{-3}, 10^{-2}, 10^{-1}\}$, such that $\Delta t(t) \leq 1$ for all times.
    (B): Speed up factor comparing the CPU times from the left panel.
    }
    \label{fig:12}
\end{figure}

As discussed in detail in \cite{auzinger2017, Koch2013}, by comparison of worker (fourth-order) and controller (third-order) we get an error estimate 
which is used to determine the next step size $\Delta t_{\rm new}$ based on the current step $\Delta t$ after each time step.
The precise rule from \cite{auzinger2017, Koch2013} is
\begin{equation}
    \Delta t_{\rm new}= \text{min}\left\{\alpha_{\text{max}}, 
    \text{max}\left[ \alpha_{\text{min}}, 
    \left(\frac{\alpha \cdot tol}{\Delta \epsilon_{\rm loc}}\right) ^ 
    {\frac{1}{\text{order} + 1}}\right]\right\} \Delta t.
    \label{eq:metric}
\end{equation}
$\Delta \epsilon_{\rm loc}$ acts as a time-local absolute error measure, and ${\rm order}$ accounts for the order of the chosen splitting scheme, in our case
fourth (w for worker) and third order (c for controller) 
\begin{equation}
 \Delta\epsilon_{\rm loc}(t) = \left(\int_{0}^{L} |\Psi_{\text{c}}(x,t) - \Psi_{\text{w}}(x,t)|^2 \,dx \right)^{\frac{1}{2}}.
 %%%\frac{\left(\int_{0}^{L} |\Psi_{\text{c}}(x,t) - \Psi_{\text{w}}(x,t)|^2 \,dx \right)^{\frac{1}{2}}}{\left(\int_{0}^{L} |\Psi_{\text{c}}(x,t)|^2 \,dx\right)^\frac{1}{2}}.
    \label{eq:L2-new}
\end{equation}
The choice of the absolute rather than the relative error, see Eq. \eqref{eq:L2}, confirmed advantageous since the latter showed a stronger dependence on the box size $L$.
Following \cite{auzinger2017}, we chose the parameters $\alpha_{\text{max}} = 4$, $\alpha_{\text{min}} = 0.25$, $\alpha = 0.9$ tested by an optimization 'by hand'.
Besides the reasonable choice $\alpha_{\rm max}>1$ and $\alpha_{\rm min}<1$, Ref. \cite{auzinger2017} states that the exact
value of the $\alpha$s is not so important. The time step is then mostly self organizing well within the mentioned limits of the parameters. 
\rev{Our simulations confirm this. What is important is to avoid too high rates in the change of the time step, which is controlled by the two parameters
$\alpha_{\text{max}}$ and $\alpha_{\text{min}}$.}
Only occasionally when the wave function $\psi$ becomes quasi-stationary, hence we get into a quasi-stable regime, large jumps of $\Delta t$ to large values are prevented by $\alpha_{\text{max}}$. \rev{On the other hand,} when two peaks in the density $|\psi|^2$ merge due to their gravitational attraction, too fast changes to small $\Delta t$ are controlled by $\alpha_{\text{min}}$. Typically for our cosmological simulations, the initial time step can be chosen relatively large, since there the mentioned problems are typically not occurring. The self adaption of the time step is quite fast, typically within just a few steps during which \rev{substantial} errors may accumulate. \rev{Our choice of} $\Delta t_{\rm ini}=10^{-3}$ turns out to be sufficient for most purposes, also in comparison with the simulation using the second-order Strang splitting. From the perspective of an end-user, getting a solution of a desired accuracy seems reasonably simple. Only the targeted error per time step, $tol$, has to be chosen, which is then compared to the local error $\Delta \epsilon_{\rm loc}$, see Eq. \eqref{eq:L2-new}. If the ratio $tol/\Delta \epsilon_{\rm loc}>1$ the time step is increased, and if $tol / \Delta \epsilon_{\rm loc}<1$ the time step is decreased in the algorithm of Ref. \cite{auzinger2017}. \edit{As mentioned after Eq. \eqref{eq:splitting}, adapting $\Delta t$ gives for each single step a unitary time evolution and hence the entire evolution remains norm preserving.}

A few results based on our adaptive-time step PS integrator have already been shown in Figs. \ref{fig:7} to \ref{fig:9}, where the conservation of mass and momentum and the energy evolution were tested. There, in all three cases, the adaptive technique compared favourably with the constant-time PS and the B-spline method.

Next, we investigate the behavior of the adaptive integrator in more detail in the following systematic manner:
Three box sizes $L\in\{100, 500, 1000\}$ are chosen as for all previous plots. To guarantee convergence in the space domain, we set the spatial grid size to the large value $N=2^{20}\approx10^6$. The initial time step is $\Delta t_{\rm ini}=10^{-3}$ for our adaptive stepper. A scan of tolerance parameters $tol\in\{10^{-1}, 10^{-2}, 10^{-3}, 10^{-4}, 10^{-5}\}$ controls the size of the time steps chosen by the rule of Eq. \eqref{eq:metric}. The relative Euclidean norm $\Delta \epsilon$, see Eq. \eqref{eq:L2}, with the adaptive-step solution and the reference solution is computed. As before, our reference were the simulations with the second-order Strang splitting at $\Delta t=\text{const.} =10^{-5}$ and the same spatial grid $N=2^{20}$. Figure~\ref{fig:11} shows these results. A second-order solution with $\Delta t=8\times10^{-5}$ serves as a guide therein. This guiding solution is systematically approached by the adaptive ones, when $tol$ is becomes smaller, see the left panel. The maximal accuracy of any solution is limited by the fidelity of the reference itself, which is estimated by $\Delta \epsilon$ of a nearby solution within the convergence plateau (see Fig.~\ref{fig:2}). Taking for this error estimate the guiding and the reference solutions, we conclude that the reference is $\mathcal{O}(10^{-5})$ accurate at the end of the simulations, i.e. at $a(t)=1$.

The right panel in Fig.~\ref{fig:11} shows how the time step evolves with time motional in the scaling parameter $a(t)$. All lines depart from the initial value $\Delta t(a_{\rm ini})=10^{-3}$.
The time step grows until $a\approx0.02$ for large $tol$. Whenever $\Delta t \geq1$, the evaluation of the step size becomes unstable and $\Delta \epsilon$ fails to serve
as a local error estimate, see data with $tol=10^{-1}$ in Fig.~\ref{fig:11}. The value of $tol$, for which this happens depends on the box size $L$, which governs the total mass in our system. When $\Delta t \leq1$, the time step is automatically reduced, typically monotonically but with possible minor fluctuations on top, e.g. at instances when the evolution changes fast due to an increased nonlinearity for instance, see at $a\approx 0.1$ in Fig.~\ref{fig:11}(B, right panel). \rev{A posteriori, after having made such detailed simulations as shown in Fig.~\ref{fig:11}(B), one may redo
the computations with the specific series of time steps, e.g. extracted by the power-law scaling of the
optimal time steps, as suggested by the figure. However, there cannot be any a priori knowledge of those
optimal values and the initial steps for which a relatively large time step is possible, see discussion above,
would have to be anyhow excluded.}

The total CPU time of our PS computation with and without adaptive time stepping is shown on the left panel in Fig.~\ref{fig:12}. All results are obtain with the hardware
reported in Sec. \ref{subsub:performance}.
In Fig.~\ref{fig:12}, every box uses the largest stable tolerance parameter $tol$ extracted from figures as the previous Fig.~\ref{fig:11} for each case independently.
An important result that we can read off Fig.~\ref{fig:12} is that all adaptive runs are much faster than the standard Strang splitting method.
By how much, is displayed on the right panel of Fig.~\ref{fig:12}, where we show the speed up factor for adaptive solutions with the highest possible tolerance
with respect to the runs with the second-order PS and constant time steps. Albeit these are not always the most precise solutions, the error $\Delta \epsilon$ is below $10^{-5}$ in all cases, for $L=500$ shown in Fig.~\ref{fig:11}. Remarkable speed up factors of $\mathcal{O}(10)$ to $\mathcal{O}(100)$ are possible, in particular for the larger spatial bases with $N \gtrsim 10^5$, where this will help most.

Three immediate applications of this time gain come to our minds: Firstly, medium-sized computations are feasible on desktop computers without massive parallelization.
Secondly, scanning a larger parameter set, e.g. of cosmological initial states or of different box sizes (impacting the total mass of the system) can be done much faster.
Lastly, it might allow us to tackle more complex problems that require larger basis sizes, also thinking of possible extensions to higher spatial dimensions.

%%%%%%%%%% Conclusion/Summary

\section{Conclusion}
\label{sec:conclusion}

We provided a detailed convergence analysis on our pseudo-spectral (PS) method for the simulation of wave-like fuzzy dark matter in
one spatial $(1+1)$ dimension(s). Our analysis is partly based on a direct comparison with an independent technique based on 
B-splines. This important comparison confirms that both methods essentially work and represent valid physical results for a problem for which
not many analytical results exist, and which is complicated by delocalised cosmological initial conditions and a time-dependent {\em and} nonlinear potential. 
In particular, detailed error measures either directly confronting two independent wave function solutions or based on conserved quantities have been studied to
underpin convergence of our results.

Unfortunately, the B-spline technique turned out to be less efficient in time and memory consumption than our PS method, typically leading to considerably larger errors at larger system sizes. A considerable increase of the basis size would be necessary for mitigating those errors which would only be possible by a massive parallelization of the method.
 
Our results show how the PS numerical simulations depend on time (CPU) and memory resources. An important advancement is the
implementation of an adaptive time step fourth-order splitting algorithm that gives a significant speed up of the order $\mathcal{O}(10)$ to $\mathcal{O}(100)$
with respect to our second-order code with fixed time steps 
\edit{without losing unitarity in the evolution}. With this speed up, we have hope to explore more complex terrain, e.g. trying to extend our approach to
$(2+1)$ dimensions. At least for small quadratic boxes this should immediately be possible, as Refs. \cite{Kopp2017, May2021, May2022, Schive2014} show us for 
two and three dimensional boxes with grid sizes of a few thousand points per direction. 
This means that memory for the spatial representation will be the limiting factor for simulations in higher dimensions. In the latter case new solutions, e.g. based on finite elements \cite{Auzinger2016} or adaptive mesh refinement \cite{Schive2014, Mina2020, Shive2018, Veltmaat2018, Niemeyer2020, Schwabe2020} should be identified, implemented, and tested in convergence and quality in full detail.

%%%%%%%%%% 
\section*{Acknowledgment}
The authors are very grateful to Luca Amendola and Massimo Pietroni for inspiring discussions and acknowledge access to the computer clusters bwHPC and HPE Apollo 4200 
through Heidelberg University and the Laboratory of scientific computation of the Universidad del Valle, respectively. 
J.M. and V.L. acknowledge support from the Colombian Science, Technology, and Innovation Fund-General Royalties System (Fondo CTeI-SGR) Contract No. BPIN 2013000100007. 
T.Z. acknowledges funding from the European Union's Horizon 2020 research and innovation  programme under the Marie Sk\l{}odowska-Curie grant agreement No. 945371.
S.W. acknowledges funding by the National Recovery and Resilience Plan (NRRP), Mission 4 Component 2 Investment 1.3 -- Call for tender No. 341 of 15/03/2022 of Italian Ministry of 
University and Research funded by the European Union -- NextGenerationEU, Project number PE0000023, Concession Decree No. 1564 of 11/10/2022 adopted by the Italian Ministry of 
University and Research, CUP D93C22000940001, Project title "National Quantum Science and Technology Institute'' (NQSTI).

%%%%%%%%%% Bibliography
%\bibliographystyle{elsarticle-num}
%\bibliography{ref} 

\end{document}